\documentclass[%
 reprint,
 amsmath,amssymb,
 aps,bm,
showpacs]{revtex4}
\usepackage{cleveref}
\usepackage[utf8]{inputenc}
\usepackage{amsmath, mathtools, tensor}
\usepackage{graphicx}
\usepackage{color}
\usepackage{dcolumn}
\usepackage{bm}
\usepackage{comment}
\usepackage{esint}
\usepackage{float}

\usepackage{graphicx} 
\usepackage{support-caption}

\usepackage[compatibility=false]{caption}

\usepackage{subcaption}

\usepackage{changepage}
\usepackage{cool}
\usepackage{xcolor}

\newcommand{\closedint}[2]{{\ensuremath{\left [ #1, #2 \right ]}}}
\newcommand{\paren}[1]{{\ensuremath{\left ( #1 \right )}}}
\newcommand{\size}[1]{{\ensuremath{\left | #1 \right |}}}
\newcommand{\sqbrk}[1]{{\ensuremath{\left [ #1 \right ]}}}

\newcommand{\map}[2]{{\ensuremath{#1 \left ( #2 \right )}}}
\newcommand{\set}[1]{{\ensuremath{\left \{ #1 \right \}}}}

\newcommand{\sgn}{\text{sgn}}

\newcommand{\DD}{{\ensuremath{\cal D}}}

\newcommand{\const}{\textrm{const}}
\newcommand{\UD}[2]{\ensuremath{^{#1}_{\phantom{#1} #2}}}
\newcommand{\UDb}[2]{\ensuremath{^{\bm #1}_{\phantom{#1} \bm #2}}}
\newcommand{\DU}[2]{\ensuremath{_{#1}^{\phantom{#1} #2}}}
\newcommand{\UDD}[3]{\ensuremath{^{#1}_{\phantom{#1} #2 #3}}}
\newcommand{\UDDD}[4]{\ensuremath{^{#1}_{\phantom{#1} #2 #3 #4}}}

\newcommand{\RR}{\ensuremath{{\mathbb{R}}}}

\newcommand{\beq}{\begin{equation}}
\newcommand{\eeq}{\end{equation}}
\newcommand{\bea}{\begin{eqnarray}}
\newcommand{\eea}{\end{eqnarray}}
\newcommand{\bean}{\begin{eqnarray*}}
\newcommand{\eean}{\end{eqnarray*}}
\newcommand{\bit}{\begin{itemize}}
\newcommand{\eit}{\end{itemize}}
\newcommand{\bfi}{\begin{figure}}
\newcommand{\efi}{\end{figure}}
\newcommand{\bfic}{\begin{figure*}}
\newcommand{\efic}{\end{figure*}}
\newcommand{\bce}{\begin{center}}
\newcommand{\ece}{\end{center}}
\newcommand{\bt}{\begin{table}}
\newcommand{\et}{\end{table}}
\newcommand{\btb}{\begin{tabular}}
\newcommand{\etb}{\end{tabular}}

\newcommand{\mba}{\mathbf A}
\newcommand{\mbb}{\mathbf B}
\newcommand{\mbc}{\mathbf C}
\newcommand{\bsmu}{\boldsymbol \mu}
\newcommand{\bsnu}{\boldsymbol \nu}

\newcommand{\calA}{\ensuremath{\mathcal{A}}}
\newcommand{\calP}{\ensuremath{\mathcal{P}}}

\newcommand{\calE}{\ensuremath{\mathcal{E}}}
\newcommand{\calO}{\ensuremath{\mathcal{O}}}

\newcommand{\calW}{\ensuremath{\mathcal{W}}}
\newcommand{\calR}{\ensuremath{\mathcal{R}}}

\newcommand{\WXL}{\ensuremath{W_{XL}}}
\newcommand{\WXX}{\ensuremath{W_{XX}}}
\newcommand{\WLX}{\ensuremath{W_{LX}}}
\newcommand{\WLL}{\ensuremath{W_{LL}}}

\newtheorem{theorem}{Theorem}[section]

\newcommand{\qed}{\nobreak \ifvmode \relax \else
      \ifdim\lastskip<1.5em \hskip-\lastskip
      \hskip1.5em plus0em minus0.5em \fi \nobreak
      \vrule height0.75em width0.5em depth0.25em\fi}

\begin{document}
\preprint{APS/123-QED}

\title{Bilocal geodesic operators in static spherically-symmetric spacetimes}
\author{Julius Serbenta}%
 \email{julius@cft.edu.pl}
\author{Miko\l{}aj Korzy\'nski} 
 \email{korzynski@cft.edu.pl}
 
\affiliation{%
 Center for Theoretical Physics, Polish Academy of Sciences,
Al. Lotnik\'ow 32/46, 02-668 Warsaw, Poland
}%

\begin{abstract}

We present a method to compute exact expressions for optical observables for static spherically symmetric spacetimes in the framework of the bilocal geodesic operator formalism. The expressions are obtained by solving the linear geodesic deviation equations for null geodesics, using the spacetime symmetries and the associated conserved quantities. We solve the equations in two different ways: by varying the geodesics with respect to their initial data and by directly integrating the equation for the geodesic deviation. The results are very general and can be applied to a variety of spacetime models and configurations of the emitter and the observer. We illustrate some of the aspects with an example of Schwarzschild spacetime, focusing on the behaviour of the angular diameter distance, the parallax distance, and the distance slip between the observer and the emitter outside the photon sphere.

\end{abstract}


\maketitle

\section{Introduction}

In general relativity, the motion of particles is affected by the spacetime geometry along their paths. Although the curvature itself cannot be observed, we can measure it directly it by studying relative motions of neighbouring freely falling particles. The key equation in this problem is the first-order, linear geodesic deviation equation (GDE), which relates relative accelerations of particles to the Riemann curvature tensor in their vicinity. Its range of application varies from tracking nearby satelites orbiting the Earth to the observation of light coming from faraway luminous bodies.

The current standard framework for optical measurements rests on gravitational lensing formalism introduced by Sachs \cite{sachs1961}. Sachs formalism was critical in theoretical GR, for example, the derivation of the Kerr metric \cite{goldberg2009, kerr1963}, with somewhat lesser importance in observational GR. It uses the geodesic deviation equation directly or in the form of optical scalar equations. The information about the influence of geometry on light is then encapsulated the expansion, shear and twist of infinitesimal bundles of rays \cite{perlick2004r}. These objects in turn can be related to various measures of distances like the angular diameter distance and the luminosity distance. Although being relatively successful in matching the observational data to the theory, the formalism is  incomplete in the following sense: it can only accommodate fixed sources and observers, so drift effects cannot be obtained from it directly \cite{korzynski2018}. Also, it is not always clear how special relativistic effects like aberration or Doppler shift affect the observables.

Recently, a new formalism of bilocal geodesic operators (BGOs) \cite{grasso2019} has been introduced, extending the previous formalism by allowing drift effects and other effects like the parallax. It is based on the resolvent of the first order GDE. As inputs, it requires the curvature along the line of sight (LOS) and the initial and final data at its endpoints. Due to their symplectic nature and the properties of null geodesics, these operators possess several symmetries. This suggests that the complete picture is simpler than it looks at first glance.

There also exist other ways to study the geometry of geodesics. One of them is the Synge's world function, which holds the information about pairs of points connected by unique geodesics. This information can be accessed by taking derivatives of the worldfunction with respect to the endpoints.
It can be shown that the second derivatives of the world function are related to the bilocal operators \cite{korzynski2021}. If the world function can be calculated exactly, the solutions of the GDE can be obtained simply by the differentiation. However, for spherically symmetric spacetimes the exact form is rarely available \cite{buchdahl1979,john1984}, and one is usually confined to a perturbative analysis of the world function and its derivatives \cite{richard1968,teyssandier2010,teyssandier2012,linet2016}.

In general, the Universe is not symmetric, which means that one has to use numerical methods to solve the propagation equations for light to obtain all optical effects. However, there are many interesting cases where the geodesic equation is integrable, and one can expect that GDE in these cases is integrable. 

In this paper we address some of these questions. We first describe the connection between the GDE and the BGOs and present their symplectic properties. Then we relate the BGOs to the variations of the geodesic with respect to its initial data and list a number of general and Killing-vector-induced conservation laws for the BGOs and the solutions of the GDE. Later, we apply all this knowledge to compute the BGOs for static spherically symmetric spacetimes and isolate physical effects by projecting our results onto the parallel propagated semi-null tetrad (SNT). In the last part of the paper, we consider the propagation of light in Schwarzschild spacetime, where we numerically investigate the behaviour of the angular diameter distance, the parallax distance, and their distance slip as we displace the emitter along the null geodesic. Finally, we reformulate these results in greater generality by studying the behaviour of BGOs in the initial, intermediate, and faraway regions.

Indeed, the problem of analytical integration of GDE is not a new one. There have been many successful attempts both for timelike \cite{fuchs1984,
jaranowski1989,fuchs1990,ellis1999} and null \cite{dwivedi1972,peters1975, peters1976, dyer1977,mlodzianowski1989,ellis1999}   geodesics, 
but the complete picture of the solutions is lacking. Often solutions assume particular initial conditions or types of orbits. Additionally, in the null case, the studies are usually limited to the behaviour of the light ray bundle projected onto the Sachs screen. This limitation completely neglects effects due to the motion of the emitter or observer.

The extension of the geometrical optics framework is also important for the present and upcoming astrophysical and cosmological observations.
For example, in the cosmological setting the parallax as well the position and redshift drifts provide additional data which can be used to study inhomogeneities and large-scale flows of matter and further constrain cosmological models \cite{korzynski2018, grasso2019}. On the other hand, the observational and computational advancements recently lead to the first images of the black hole shadow \cite{event2019}. The theory behind it is well-developed \cite{perlick2021}, but not entirely complete. In these problems the observer is usually considered to be static or comoving with some global flow. It would be interesting to see whether the BGO formalism could be used to make the problem fully covariant and reveal new properties of the black hole shadow. Static spherically symmetric spacetimes are good starting points for such studies because they are sufficiently simple while still being good models for various types of massive compact objects.

\paragraph{Applications}
Due to assumed symmetries, all possible applications concerning will be limited to static spacetimes with spherical symmetries. Killing vectors allow us to integrate equations exactly, and solutions include only a handful of integrals of functions of metric coefficients along the trajectory of light. Furthermore, observer effects like stellar aberration with the arbitrary alignment of observer's four-velocity are taken into account by appropriate parallel transports. Moreover, general treatment of geodesic deviation allows us to characterize the formation of caustics in a more precise manner. Now we can state precise conditions for the formation of focal or conjugate points in terms of parameters of the null geodesic. Similarly, we can quantify how the size and the shape
of an image as seen by the observer depends on the positions of emitters and observers. The formalism applied here treats all optical effects on the same footing, so this information is related to previously mentioned effects and forms a consistency requirement between all of them.

In practise this means that we can study cases when the lensing and lensed structures do not fit the traditional lensing formalism, e.g. when the impact parameter of the light or distances between emitters, observers and lensing bodies are not much larger than Schwarzschild radius. Geodesic bilocal operator formalism holds both in weak and strong lensing regimes as well as all intermediate cases. Hence, we are able to patch these results and explain transitions from one regime to the other one. It is worth mentioning that General Relativity is, in general, not assumed here. We only require a 4-dimensional Lorentzian metric theory of gravity. The conclusions about the spacetime we reach are purely geometric. Thus, the physical interpretation depends on the choice of the theory of gravity. 
\paragraph{Structure of the paper}

In Sec. II, we begin with formulating bilocal geodesic operators (BGOs) in the geometric optics regime and restating some of their properties. Then we sketch one of the methods of calculating them, based on  the variation of null geodesic with respect to initial data. The second method employs Killing conservation to reduce GDE to a system of coupled first order ordinary differential equations (ODEs), which we integrate, and is described in Sec. III. In Sec. III we also find expressions of optical observables for the emitter and observer travelling arbitrarily and describe their behaviour, with detailed derivation given in the appendices. In Sec. IV we estimate effects for a Schwarzschild black hole for static observers and emitters. We state our conclusions in Sec. V.

\paragraph{Notation}
Greek letters $\paren{\alpha,\beta,\dots}$ run from $0$ to $3$, and uppercase Latin indices run from $1$ to $2$. They all enumerate tensor components in the coordinate tetrad. In some rare cases the uppercase Latin indices are also used to label linearly independent solutions of differential equations. Boldface versions of indices cover the same range but denote components in the SNT, defined in Section \ref{sec:parallel}, as opposed to the coordinate tetrad. The dot denotes the derivative with respect to the affine parameter along the null geodesic. Prime denotes differentiation with respect to $r$. Subscript $\mathcal O$ and $\mathcal E$ denote evaluation of the quantity at respectively the point of observation and emission, i.e. $f_{\mathcal {O}} \equiv \map f {\lambda_{\mathcal O}}$. 

We introduce the following short-hand notation for integrals over a null geodesic, performed both over the affine parameter and the radial coordinate $r$. These integrals have 
 common kernels which we will denote $\paren{\ell^r}^{-2}$ or $\paren{\ell^r}^{-3}$ as well as a varying part composed of the metric coefficients $A(r)$, $B(r)$ and $C(r)$. Namely:
\bea
\begin{split}
I_B &= \int_0^\lambda \dfrac{d \tilde \lambda}{ \map B {\map r \lambda} \paren{\ell^r}^2} = \fint_{r_\calO}^{r_\calE} \dfrac{d \tilde r}{\map B {\tilde r} \paren{\ell^r}^3}\\
I_{AB} &= \int_0^\lambda \dfrac{d \tilde \lambda}{\map A {\map r \lambda} \map B {\map r \lambda} \paren{\ell^r}^2} = \fint_{r_\calO}^{r_\calE} \dfrac{d \tilde r}{\map A {\tilde r} \map B {\tilde r} \paren{\ell^r}^3}\\
I_{BC} &= \int_0^\lambda \dfrac{d \tilde \lambda}{\map B {\map r \lambda} \map C {\map r \lambda} \paren{\ell^r}^2} = \fint_{r_\calO}^{r_\calE} \dfrac{d \tilde r}{\map B {\tilde r} \map C {\tilde r} \paren{\ell^r}^3}\\
I_{ABC} &= \int_0^\lambda \dfrac{d \tilde \lambda}{\map A {\map r \lambda} \map B {\map r \lambda} \map C {\map r \lambda} \paren{\ell^r}^2} = \fint_{r_\calO}^{r_\calE} \dfrac{d \tilde r}{\map A {\tilde r} \map B {\tilde r} \map C {\tilde r} \paren{\ell^r}^3}\\
\end{split} 
\eea
The slash  reminds us that in the case of a turning point along the photon path, the integral over $r$ must be split into segments with appropriately chosen signs of the integrand, see \cite{kapec2019}. 

We assume the speed of light $c = 1$.

\section{Formulation}
Let $\mathcal M$ be a smooth Lorentzian manifold with a metric $g$ of signature $\paren{-, +, +, +}$. Let $\paren{\zeta^\mu}$ be a coordinate system. Let $\gamma : 
\closedint {\lambda_\calO} \lambda \to \mathcal M$ be a geodesic connecting two points, $x_\calO$ and $x_\calE$, with affine parameter values $\lambda_\calO$ and $\lambda$ respectively. We also introduce two tetrads for decomposing geometric objects:  $\paren{\partial_\mu}$ will denote the coordinate tetrad associated with $\paren{\zeta^\mu}$, while $\paren{e_{\boldsymbol \mu}}$ will denote the tetrad, which is parallel transported along $\gamma$.  

We choose a coordinate system which covers the neighborhoods of both endpoints of $\gamma$. Then the geodesic curve $\map {x^\mu} {x_\calO, \ell_\calO, \lambda}$ is a function of the initial point $x_\calO$, the initial tangent vector $\ell_\calO$, and the value of the affine parameter $\lambda$, corresponding to the geodesic with aforementioned initial conditions.

Now we perturb the initial data of the geodesic at $\lambda_\calO$ according to $x_\calO^\mu \to x_\calO^\mu + \delta x_\calO^\mu$, $\ell_\calO^\mu \to \ell_\calO^\mu + \delta\ell_\calO^\mu$ in
a coordinate tetrad. Up to the linear order in perturbation, the deviation vector $\delta x^\mu = \xi^\mu$ satisfies the following first order GDE \cite{levi1925,synge1927}:
\bea
\label{eq:gde}
\nabla_\ell \nabla_\ell \xi^\mu - R \UDDD \mu \ell \ell \nu \xi^\nu = 0
\eea
In the literature the tensor
\bea
\calR \UD \mu \nu = R \UDDD \mu \alpha \beta \nu \ell^\alpha \ell^\beta
\label{eq:otm}
\eea
is also known as the (optical) tidal matrix or optical tidal tensor.

 The deviation at a different point, corresponding to a different value of $\lambda$, will take the following form:
\begin{equation}
\begin{split}
    \delta x^\mu &= {\WXX}\UD \mu \nu \, \delta x^\nu_\calO + {\WXL} \UD {\mu} {\nu} \,\Delta \ell^\nu_\calO\\
    \Delta \ell^\mu &= {\WLX} \UD \mu \nu \, \delta x^\nu_\calO + {\WLL} \UD \mu \nu \,\Delta \ell^\nu_\calO
\end{split} \label{eq:W1}
\end{equation}
where  $\delta x^\mu_\calO$, $\delta x^\mu$ are the position perturbations and $\Delta \ell^\mu_\calO$, $\Delta \ell^\mu$ are the covariant perturbations of the tangent vector at $\lambda_\calO$ and $\lambda$ respectively. 
The covariant perturbations of tangent vectors are defined by
\begin{equation}
\label{eq:cdif}
    \begin{split}
        \Delta \ell^\mu_\calO &= \delta \ell^\mu_\calO + \map {{\Gamma}\UDD \mu \alpha \beta} {x_\calO} \,\ell^\alpha_\calO \,\delta x^\beta_\calO\\
        \Delta \ell^\mu &= \delta \ell^\mu + \map {{\Gamma}\UDD \mu \alpha \beta} {x}\, \ell^\alpha \,\delta x^\beta.
    \end{split}
\end{equation}
The Eqs. \eqref{eq:gde} and \eqref{eq:W1} are related by
\bea
\begin{split}
\map {\xi^\mu} {\lambda_\calO} &= \delta x^\mu_\calO\\
\map {\xi^\mu} {\lambda_\calE} &= \delta x^\mu\\
\map {\nabla_\ell \xi^\mu} {\lambda_\calO} &= \Delta \ell^\mu_\calO\\
\map {\nabla_\ell \xi^\mu} {\lambda_\calE} &= \Delta \ell^\mu.
\end{split}
\eea 
Here  $\WXX, \WXL, \WLX, \WLL$ are bitensors mapping tangent vectors from $\calO$ to $\calE$. Together they form the bilocal geodesic operator $\calW: T_\calO M \oplus T_\calO M \mapsto T_\calE M \oplus T_\calE M$, defined by the linear relation
\bea
\begin{pmatrix}
\delta x^\mu \\ \Delta \ell^\nu
\end{pmatrix} = \calW \,\begin{pmatrix}
\delta x_\calO^\alpha \\ \Delta \ell_\calO^\beta\end{pmatrix}.
\eea
$\calW$ and its four constituent bitensors may be expressed as functionals of the Riemann
curvature tensor along the line of sight. 
Namely, $\calW$ expressed in a parallel-propagated tetrad plays the role of the resolvent of the GDE
with $\calO$ as the starting point and therefore satisfies the resolvent ODE when expressed in the parallel propagated tetrad. 
In the same way, four bitensors 
can be expressed as solutions of 
appropriate matrix ODEs written in any parallel propagated tetrad
\cite{grasso2019}.
Namely, consider the matrix ODE
\begin{equation}
\label{eq:gdeM}
    \dfrac{d^2}{d \lambda^2}{A}\UD{\boldsymbol \mu}{\boldsymbol \nu} - \calR \UD {\boldsymbol \mu} {\boldsymbol \sigma} \, {A} \UD {\boldsymbol \sigma}{\boldsymbol \nu} = 0.
\end{equation}
This equation has to be supplied with the initial conditions. Suppose \bea \label{eq:gdeic1}
 \map{A\UD{\boldsymbol \mu}{\boldsymbol \nu}}{\lambda_\calO} &=&
 \delta \UD{\boldsymbol \mu}{\boldsymbol \nu} \\
  \dfrac{d}{d \lambda}
 \map{A\UD{\boldsymbol \mu}{\boldsymbol \nu}}{\lambda_\calO} &=&
 0.
\eea
Then $W_{XX}$ and $W_{LX}$ are given by

\bea
{W_{XX}} \UD{\boldsymbol \mu}{\boldsymbol \nu} &=& A\UD{\boldsymbol \mu}{\boldsymbol \nu}(\lambda) \\
 {W_{LX}} \UD{\boldsymbol \mu}{\boldsymbol \nu} &=& \dfrac{d}{d\lambda}A\UD{\boldsymbol \mu}{\boldsymbol \nu}(\lambda)
\eea
In a similar way, for $\WLX$ and $\WLL$ we have
\bea
 \map{A\UD{\boldsymbol \mu}{\boldsymbol \nu}}{\lambda_\calO} &=& 0
  \\
  \dfrac{d}{d \lambda}
 \map{A\UD{\boldsymbol \mu}{\boldsymbol \nu}}{\lambda_\calO} &=&
 \delta \UD{\boldsymbol \mu}{\boldsymbol \nu} \\
 {W_{XL}} \UD{\boldsymbol \mu}{\boldsymbol \nu} &=& A\UD{\boldsymbol \mu}{\boldsymbol \nu}(\lambda) \\
 {W_{LL}} \UD{\boldsymbol \mu}{\boldsymbol \nu} &=& \dfrac{d}{d\lambda}A\UD{\boldsymbol \mu}{\boldsymbol \nu}(\lambda)  \label{eq:gdeic2}.
\eea
$\calW$ and the constituent bitensors are not arbitrary. Irrespective of spacetime geometry, they need to satisfy several algebraic conditions. Firstly,
 $\calW$ is always a symplectic mapping in the following sense: consider the matrix $\calW$ defined by
 \begin{equation}
     \calW = \paren {\begin{array}{cc}
         {W_{XX}}\UD \mu \alpha & {W_{XL}} \UD \mu \beta \\
         {W_{LX}}\UD \nu \alpha & {W_{LL}} \UD \nu \beta
     \end{array}}
 \end{equation}
 and the nondegenerate, antisymmetric matrix
 \bea
 \Omega = \paren{\begin{array}{cc}
         0 & g_{\alpha \beta} \\
         -g_{\gamma \delta} & 0
     \end{array}}.
 \eea
 Then we have
 \begin{equation}
    \label{eq:symp}
     \calW^T \Omega\, \calW = \Omega, 
 \end{equation}
 where
 \begin{equation}
     \calW^T = \paren{\begin{array}{cc}
         \paren{{W_{XX}}^T}\DU \alpha \mu & \paren{{W_{LX}}^T} \DU \alpha \nu \\
         \paren{{W_{XL}}^T} \DU \beta \mu & \paren{{W_{LL}}^T} \DU \beta \nu
     \end{array}}
 \end{equation}
and ${W_{**}} \UD \mu \alpha = \paren {W_{**}^T} \DU \alpha \mu$, see \cite{uzun2020}. The transpose in \eqref{eq:symp} is the usual matrix transpose. The transpose of the BGOs changes the order of their tensorial indices without spoiling contractions with other terms in tangent spaces at both $\calO$ and $\calE$. The proof of $\eqref{eq:symp}$ is relatively simple: we begin by recalling the ODE for $\calW$ and its initial data. In a parallel propagated tetrad we have
 \begin{eqnarray*}
     \map \calW \calO &=& \paren {\begin{array}{cc}
         \delta \UD {\boldsymbol \mu} {\boldsymbol \alpha} & 0  \\
         0 & \delta \UD {\boldsymbol \nu} {\boldsymbol \beta} 
     \end{array} } \\
     \dfrac d {d\lambda} \calW &=& S\,  \calW
 \end{eqnarray*}
 with
 \bea
  S = \paren{ \begin{array}{cc}
         0 & \delta \UD {\boldsymbol \mu} {\boldsymbol \beta} \\
         R \UDDD {\boldsymbol \nu} \ell \ell {\boldsymbol \alpha} & 0
     \end{array}}.
 \eea
By taking its transpose we get
 \begin{equation}
     \dfrac d {d\lambda} \calW^T = \calW^T\,S^T .
 \end{equation}
Taking the derivative of \eqref{eq:symp} gives now
\bea
\frac{d}{d\lambda}\,\left(\calW^T \,\Omega\, \calW\right) = \calW^T (S^T\, \Omega + \Omega \,S)\calW.
\eea
The term $S^T\, \Omega + \Omega \,S$  inside the brackets vanishes due to the symmetry of the Riemann tensor $R_{\alpha \ell\ell\beta} = R_{\beta \ell\ell \alpha}$, so the 
left-hand side of \eqref{eq:symp} is constant. We can also use the initial condition for $\calW$ at $\calO$ to show  that \eqref{eq:symp} holds at $\calO$ and thus everywhere along $\gamma_0$.

Secondly, we note here  additional relations involving the tangent vectors to the LOS $\ell_\calO$ and $\ell$ at $\lambda_\calO$ and $\lambda$ respectively \cite{grasso2019}:
\begin{equation}\label{eq:additionalrelations}
    \begin{split}
        {W_{XX}}\UD \mu \nu \ell^\nu_\calO & = \ell^\mu_\calE\\
        {W_{XL}}\UD \mu \nu \ell^\nu_\calO & = \paren{\lambda_\calE - \lambda_\calO} \ell^\mu_\calE\\
        {W_{LX}}\UD \mu \nu \ell^\nu_\calO & = 0\\
        {W_{LL}}\UD \mu \nu \ell^\nu_\calO & = \ell^\mu_\calE\\ \\
        \ell_{\calE\mu} {W_{XX}}\UD \mu \nu & = \ell_{\calO \nu}\\
        \ell_{\calE\mu} {W_{XL}}\UD \mu \nu  & = \paren{\lambda_\calE - \lambda_\calO} \ell_{\calO \nu}\\
        \ell_{\calE\mu} {W_{LX}}\UD \mu \nu  & = 0\\
        \ell_{\calE\mu} {W_{LL}}\UD \mu \nu  & = \ell_{\calO \nu}
    \end{split}
\end{equation}

\subsection{Bilocal geodesic operators from the variations of the general solution of the geodesic equation} \label{sec:solution}

The set of equations \eqref{eq:gdeM} and \eqref{eq:gdeic1}-\eqref{eq:gdeic2} constitutes a system of second order ODEs. Its solution can be found analytically only in the simplest cases. However, it turns out that it is possible to circumvent this problem if we know the general solution of the geodesic equation on our manifold in an explicit or implicit form. In that case the components of $\calW$ can be found by simple differentiation. This approach is not new and has been considered previously \cite{bazanski1989,jaranowski1989}, but only in the context of the Hamilton-Jacobi equation for the geodesic motion: suppose we have the solution to the geodesic equation expressed in terms of the curve parameter and the integration constants. Suppose also that we have a complete integral of the associated Hamilton-Jacobi equation. Then the variation of this integral with respect to the coordinates of the geodesic and the geodesic constants yields the solution to the GDE. 

The method we present is a bit different. It avoids the Hamilton-Jacobi equation and provides a direct path from the solution of the geodesic equation to the BGOs. The technique that we will apply has already been described partially in \cite{korzynski2020}, i.e. without the operators $W_{LX}$ and $W_{LL}$, and
we will describe it now in full detail. It is coordinate-dependent in the sense that it requires fixing a coordinate system in which we know how to solve the geodesic equation. 

Let  $\map {x^\mu} {x^\nu_\calO, \ell^\nu_\calO, \lambda}$ be the general solution to the geodesic equation written in  coordinates $(\xi^\mu)$ with the initial data
$x^\mu(\lambda_\calO) = x_\calO^\mu$, $\ell^\mu(\lambda_\calO) = \ell_\calO^\mu$. Let
$x_\calE^\mu$ denote the coordinates of
the second endpoint of the geodesic, corresponding to a fixed value
of the affine parameter
$\lambda = \lambda_\calE$, and let $\ell_\calE^\mu$ denote the tangent vector to the geodesic in $\lambda = \lambda_\calE$. 
Let us now consider the \emph{full covariant} variation with respect to \emph{all} variables, including $\lambda$, of the
second endpoint  $\calE$ ,
taken at $(x_\calO^\mu,\ell_\calO^\mu, \lambda_\calE)$. It reads:
\bea
        \delta x^\mu_\calE &=& {\WXX}\UD \mu \nu \, \delta x^\nu_\calO + {\WXL}\UD \mu \nu \, \Delta \ell^\nu_\calO + \ell^\mu_\calE \,\delta \lambda \label{eq:var01}\\ 
        \Delta \ell^\mu_\calE &=& {\WLX}\UD \mu \nu \, \delta x^\nu_\calO + {\WLL}\UD \mu \nu\, \Delta \ell^\nu_\calO .
        \label{eq:var02}
\eea
These relations generalize (\ref{eq:W1}) to the situation when we allow the affine parameter of the second endpoint of the geodesic to vary as well, i.e. $\lambda = \lambda_\calE + \delta \lambda$. They follow simply from (\ref{eq:W1}) and the definition of a geodesic. Namely, the position variation under the variations of $\lambda$, with fixed initial data ($\delta x_\calO^\mu = \Delta l_\calO^\mu = 0$), is by definition given by the tangent vector
$\ell^\mu_\calE$, and hence the last term in (\ref{eq:var01}). On the other hand, the \emph{covariant} variations of the tangent vector along any fixed geodesic $\gamma$ must vanish because the tangent vector $\ell^\mu$ is covariantly constant ($\nabla_\ell \ell^\mu = 0$), and hence no $\delta\lambda$ term
in (\ref{eq:var02}).

Now, it follows from Eqs. \eqref{eq:var01} and \eqref{eq:var02} that by taking the full solution to the geodesic equation in a given coordinate system $(\xi^\mu)$, differentiating it with respect to all the components of $x^\mu_\calO, \ell^\mu_\calO$ and $\lambda$, and expressing the results in terms of covariant differentials, component by component, one can recover all the BGOs. Their components, expressed in the coordinate tetrads of the coordinate system $(\xi^\mu)$, play simply the role of the expansion coefficients in the basis $(\delta x_\calO^\mu, \Delta \ell_\calO^\mu, \delta\lambda)$. With this result in hand, we are now ready to  describe step by step how we can evaluate $\cal W$ from the derivatives of
the general solution of the geodesic equation.

We begin with ordinary total variations of $\map {x^\mu} {x^\nu_\calO, \ell^\nu_\calO, \lambda}$ and the tangent vector
$\displaystyle \ell^\mu(x^\nu_\calO, \ell^\nu_\calO, \lambda) = \frac{\partial x^\mu}{\partial \lambda}$, taken at $\lambda = \lambda_\calE$:
\begin{equation}
\begin{split} \label{eq:variations}
    \delta x^\mu_\calE &= \paren{\dfrac{\partial x^\mu_\calE}{\partial x^\nu_\calO}}_{\ell_\calO, \lambda} \delta x^\nu_\calO + \paren{\dfrac{\partial x^\mu_\calE}{\partial \ell^\nu_\calO}}_{x_\calO, \lambda} \delta \ell^\nu_\calO + \paren{\dfrac{\partial x^\mu_\calE}{\partial \lambda}}_{x_\calO, \ell_\calO} \,\delta \lambda\\
    \delta \ell^\mu_\calE &= \paren{\dfrac{\partial \ell^\mu_\calE}{\partial x^\nu_\calO}}_{\ell_\calO, \lambda} \delta x^\nu_\calO + \paren{\dfrac{\partial \ell^\mu_\calE}{\partial \ell^\nu_\calO}}_{x_\calO, \lambda} \delta \ell^\nu_\calO + \paren{\dfrac{\partial \ell^\mu_\calE}{\partial \lambda}}_{x_\calO, \ell_\calO} \delta \lambda 
\end{split}
\end{equation}
Just like in thermodynamics, the subscripts denote here variables kept fixed during respective variations. Note also that we have used $\delta x_\calE^\mu$ for the variation of $x^\mu$ and $\delta \ell_\calE^\mu$ for the variation of $\ell^\mu$. Now we apply \eqref{eq:cdif} to change the basis of variations from $\paren {\delta x^\mu_\calO, \delta \ell^\mu_\calO, \lambda}$ to $\paren {\delta x^\mu_\calO, \Delta \ell^\mu_\calO, \lambda}$ and switch from $\delta l_\calE^\mu$ to $\Delta l_\calE^\mu$ in the second equation.
 Together with \eqref{eq:var01} and \eqref{eq:var02}, this leads to the following relations:
\begin{equation}
\label{eq:covW}
    \begin{split}
       & {\WXL}\UD \mu \nu = \paren{\dfrac{\partial x^\mu_\calE}{\partial \ell^\nu_\calO}}_{x_\calO, \lambda} \quad \\
       &{\WXX} \UD \mu \nu = \paren{\dfrac{\partial x^\mu_\calE}{\partial x^\nu_\calO}}_{\ell_\calO, \lambda} - {\WXL}\UD \mu \beta \map {\Gamma \UDD \beta \alpha \nu} {x_\calO} \,\ell^\alpha_\calO\\
       & {\WLL} \UD \mu \nu = \paren{\dfrac{\partial \ell^\mu_\calE}{\partial \ell^\nu_\calO}}_{x_\calO, \lambda} + \map {\Gamma \UDD \mu \alpha \beta} {x_\calE} \,\ell^\alpha_\calE\, {\WXL}\UD \beta \nu\\
       & {\WLX} \UD \mu \nu = \paren{\dfrac{\partial \ell^\mu_\calE}{\partial x^\nu_\calO}}_{\ell_\calO, \lambda} + \map {\Gamma \UDD \mu \alpha \beta} {x_\calE}\, \ell^\alpha_\calE\, {\WXX} \UD \beta \nu - {\WLL} \UD \mu \beta \map {\Gamma \UDD \beta \alpha \nu} {x_\calO}\, \ell^\alpha_\calO + \map {\Gamma \UDD \mu \alpha \gamma}{x_\calE}\, \ell^\alpha_\calE\, {{\WXL} \UD \gamma \beta} \map {\Gamma \UDD \beta \alpha \nu} {x_\calO}\, \ell^\alpha_\calO
    \end{split}
\end{equation}
We have thus expressed the four bitensors by $\ell_\calO^\mu$, the Christoffel symbols at $\calO$ and $\calE$ and the derivatives of $x^\mu(x_\calO^\nu,\ell_\calO^\nu,\lambda)$ (the first derivatives 
$\displaystyle \frac{\partial x^\mu_\calE}{\partial x^\nu_\calO}, \frac{\partial x^\mu_\calE}{\partial \ell_\calO^\nu}, \ell_\calE^\mu \equiv \frac{\partial x_\calE^\mu}{\partial \lambda}$ and the
second derivatives $\displaystyle \frac{\partial \ell_\calE^\mu}{\partial x_\calO^\nu}  \equiv
\frac{\partial^2 x_\calE^\mu}{\partial \lambda \partial x_\calO^\nu}, \frac{\partial \ell_\calE^\mu}{\partial \ell_\calO^\nu} \equiv
\frac{\partial^2 x_\calE^\mu}{\partial \lambda \partial \ell_\calO^\nu}$). 
These  bitensors are sufficient to reconstruct optical observables, such as the matrix of magnification and parallax, as well as position and redshift drifts (see \cite{grasso2019}).

When we apply the BGO formalism to light rays, we need to impose one more requirement for the variations of the endpoints. Namely, we limit the admissible variations to those which, at the leading, linear order, preserve the null character of the corresponding geodesics:
\begin{equation}
    \Delta \ell^\sigma_\calO\, \ell_{\calO \sigma} = 0. \label{eq:nullvariations}
\end{equation}
This means that not all variations of $\ell_\calO$ are allowed, and therefore in practice we only consider $\cal W$ restricted to a subspace of codimension 1, defined by the condition (\ref{eq:nullvariations}).
Still, as has been shown in \cite{korzynski2020}, the reduced operator defined on this subspace carries sufficient information 
to calculate the optical observables. 
In this paper, however, we assume that a geodesic can be found for an arbitrary causal character. We thus denote the normalization parameter by $\epsilon$, and allow it to vary arbitrarily:
\begin{equation}
    \Delta \ell^\sigma_\calO \ell_{\calO \sigma} = \delta \epsilon.
\end{equation}

The method described above works if we have explicit expressions for all the components of the geodesic. It may happen, however, that the full (or part of) solution can be found only formally, i.e. involving the inverse of a non-elementary function, often given as a quadrature. In such circumstances the method still works with a  slight modification. Suppose we have a system of 8 equations of the form $\map {f_i} {x_\calE^\mu, \ell_\calE^\mu, x_\calO^\mu, \ell_\calO^\mu, \lambda} = 0$ defining the geodesics implicitly. We may compute the total differential of each relation:
\bea
0 = \frac{\partial f_i}{\partial x_\calE^\mu} \,\delta x_\calE^\mu + \frac{\partial f_i}{\partial \ell_\calE^\mu} \,\delta \ell_\calE^\mu +
\frac{\partial f_i}{\partial x_\calO^\mu} \,\delta x_\calO^\mu  +\frac{\partial f_i}{\partial \ell_\calO^\mu} \,\delta \ell_\calO^\mu  +
\frac{\partial f_i}{\partial \lambda}\,\delta \lambda.
\label{eq:relations}
\eea
The next step is to solve this system of linear relations between the differentials for the components $\delta x_\calE^\mu$, $\delta \ell_\calE^\mu$. 
In this step we are working with linear relations only, so the operation is conceptually simple, though potentially tedious, and should always work given a sufficient number of linearly independent relations (\ref{eq:relations}).
In this way we obtain eight equations 
in which both $\delta x_\calE^\mu$ and $\delta \ell_\calE^\mu$ are expressed as linear combinations of $\delta x_\calO^\mu, \delta \ell_\calO^\mu$ and $\delta \lambda$. We have thus transformed the linearized relations (\ref{eq:relations}) to a form consistent with equation (\ref{eq:variations}). 
In the final step, as before, we pass from $\delta \ell_\calO^\mu$ and $\delta \ell_\calE^\mu$ to their covariant
counterparts $\Delta \ell_\calO^\mu$ and $\Delta \ell_\calE^\mu$ using
\eqref{eq:cdif}. This way we  transform our relation to the form (\ref{eq:var01})-(\ref{eq:var02}), from which we may read off directly all 
components of $\cal W$ expressed in the coordinate tetrads of the coordinate system $(\xi^\mu)$.

\subsection{Bilocal geodesic operators and Killing vectors}

In a spacetime admitting a Killing vector we may derive additional identities and algebraic relations for the BGO, which enormously simplify the problem of determining the components of the BGO in a given spacetime.

Assume that the spacetime admits a Killing vector $K^\mu$, i.e. $\nabla_\mu K_\nu + \nabla_\nu K_\mu = 0$. 
Since the flow of $K^\mu$ is an isometry, it must also map geodesics into geodesics. Therefore, the deviation of the fiducial null geodesic by $K^\mu$ must preserve the geodesic character of the curve at linear and other orders.
It follows that $\xi^\mu(\lambda) = K^\mu$ must be a valid solution of the GDE \cite{manoff1979}.
It can also be proved by direct differentiation of the Killing condition that
$\nabla_\ell \nabla_\ell K^\mu - R\UD{\mu}{\nu\alpha\beta}\,\ell^\nu\,\ell^\alpha\,K^\beta = 0$.
This means that the initial data of the GDE at $\calO$ of the form
$\delta x_\calO^\mu = K_\calO^\mu$, $\Delta \ell_\calO^\mu = \paren{ \nabla_\ell K^\mu}_\calO$ must be mapped into the initial data at $\calE$
of the form $\delta x_\calE^\mu = K_\calE^\mu$, $\Delta \ell_\calE^\mu = \paren{\nabla_\ell K^\mu}_\calE$. From (\ref{eq:W1}) applied to $\calO$ and $\calE$ we obtain then the following identities:
\bea
K_\calE^\mu &=& {W_{XX}}\UD{\mu}{\nu}\,K_\calO^\nu +
{W_{XL}}\UD{\mu}{\nu}\,\left(\nabla_\kappa K^\nu\right)_\calO\,\ell_\calO^\kappa \label{eq:WvsK1}\\
\left(\nabla_\kappa K^\mu\right)_\calE\,\ell_\calE^\kappa&=& {W_{LX}}\UD{\mu}{\nu}\,K_\calO^\nu +
{W_{LL}}\UD{\mu}{\nu}\,\left(\nabla_\kappa K^\nu\right)_\calO\,\ell_\calO^\kappa \label{eq:WvsK2}
\eea
A dual set of  identities can be derived by combining (\ref{eq:WvsK1})-(\ref{eq:WvsK2}) with the symplectic condition (\ref{eq:symp}):
\bea
K_{\calO\,\mu} &=& K_{\calE\,\nu}\,{W_{LL}}\UD \nu \mu - (\nabla_\sigma K_\nu)_\calE\,\ell_{\calE}^\sigma\,{W_{XL}}\UD \nu \mu \label{eq:WvsKdual1}\\
(\nabla_\sigma K_\mu)_\calO \,\ell_\calO^\sigma &=& -K_{\calE\,\nu}\,{W_{LX}}\UD \nu \mu + (\nabla_\sigma K_\nu)_\calE\,\ell_{\calE}^\sigma\,{W_{XX}}\UD \nu \mu  \label{eq:WvsKdual2}.
\eea
This can be shown by inverting (\ref{eq:WvsK1})-(\ref{eq:WvsK2}) with the help of $\calW^{-1}$ and using (\ref{eq:symp}) to rewrite $\calW^{-1}$ in terms of $\calW^T$ and $\Omega$.

\subsection{The linear GDE and its conserved quantitites}

The existence of Killing vectors also affects the properties of the  GDE and its solutions. Namely,  suppose that $\xi^\mu$ satisfies the GDE along the geodesic, and $K^\mu$ is a Killing vector. Then the following quantity is conserved along the geodesic curve \cite{fuchs1977}:
\begin{equation}
\label{eq:gdek}
    \xi_\mu \nabla_\ell K^\mu - K_\mu \nabla_\ell \xi^\mu = \Sigma
\end{equation}

This can be proven in the following way. We have that both $\xi$ and $K$ satisfy the GDE:
\begin{equation}
\begin{split}
\nabla_\ell \nabla_\ell \xi^\mu - \calR \UD \mu \nu \xi^\nu &= 0\\
\nabla_\ell \nabla_\ell K^\mu - \calR \UD \mu \nu K^\nu &= 0.
\end{split}
\end{equation}
Now contract the first equation with $K_\mu$ and the second one with $\xi_\mu$, and subtract one from the other. Due to the symmetries of Riemann tensor we are left with
\bea
K_\mu \nabla_\ell \nabla_\ell \xi^\mu - \xi_\mu \nabla_\ell \nabla_\ell K^\mu = 0.
\eea
Finally, use the linearity of the covariant derivative and cancel out similar terms to write the expression as a covariant derivative along $\ell$:
\bea
\nabla_\ell \paren {K_\mu \nabla_\ell \xi^\mu - \xi_\mu \nabla_\ell K^\mu} = 0.
\eea
We can also assign a physical meaning to the quantity $\Sigma$ \cite{mlodzianowski1989}. Suppose $C$ is a conserved quantity generated by a Killing vector: $C = K^\mu \ell_\mu$. Now let us take a covariant derivative along the deviation vector $\xi$. Recalling that $\xi$ is Lie dragged along $\ell$ and that $K$ is a Killing vector, we can show that:
\begin{equation}
    \nabla_\xi C = \Sigma
\end{equation}
Hence, $\Sigma$ is simply a variation of a Killing conserved quantity along $\xi$. 

Apart from the conservation laws connected with the Killing vectors, we automatically have two conserved quantities in GDE in any spacetime. Let the geodesic tangent vector $\ell^\mu$ and the deviation vector $\xi^\mu$ be evaluated at the same point $\map {x^\mu} \lambda$ along the geodesic. Then:
\begin{equation}
\label{eq:inner}
    \ell^\mu \xi_\mu =  \mathcal A + \mathcal B \lambda
\end{equation}
where $\mathcal A$ and $\mathcal B$ are constants. Thus we have the conservation of two quantities: 
$\mathcal B = \nabla_\ell \xi^\mu\,\ell_\mu$ and $\mathcal A = \ell^\mu \xi_\mu - \lambda\cdot \nabla_\ell \xi^\mu\,\ell_\mu$.
Finally, due to the symmetries of Riemann tensor, any two solutions $\xi_1$, $\xi_2$ of the GDE generate a constant of integration:
\bea
\xi_{1\mu} \nabla_\ell {\xi_2}^\mu - {\xi_2}_\mu \nabla_\ell {\xi_1}^\mu = \const.
\eea 
This  expression is bilinear and antisymmetric in the solutions $\xi_1$ and $\xi_2$. It defines a conserved symplectic form in the space of solutions \cite{uzun2020}.

\section{Static spherically symmetric spacetimes}

\subsection{Solution of the geodesic equation in arbitrary spherical coordinates}

A static spherically symmetric spacetime has the metric of the following form:
\begin{equation}
    g=-\map A r dt^2+ \map B r dr^2+\map C r \paren{d\theta^2+\sin{\theta}^2d\phi^2}.
\end{equation}
The radial coordinate may be reparametrized $r \to r(\tilde r)$. In particular, without the loss of generality, we may choose the area radius $r$, defined by $C(r) = r^2$. Nevertheless,  in the subsequent calculations we will keep the radial coordinate general with $A(r), B(r), C(r)$ treated as independent functions.

The spacetime has a 4-parameter symmetry group generated by four Killing vectors  \cite{eiesland1925}:
\begin{equation}
\begin{split}
    &T^\mu=\paren{1,0,0,0}\\
    &\Phi_x^\mu=\paren{0,0,-\sin{\phi},-\cot{\theta}\cos{\phi}}\\
    &\Phi_y^\mu=\paren{0,0,\cos{\phi},-\cot{\theta}\sin{\phi}}\\
    &\Phi_z^\mu=\paren{0,0,0,1}
\end{split}
\end{equation}
We will now briefly present the derivation of the general solution of the geodesic equation in the implicit form. This is a well-known material, but we present it here for completeness and to introduce the notation for the following sections.

 The Killing vectors simplify the problem of solving the geodesic equation. Namely, each Killing vector generates a conserved quantity along the geodesic:
\begin{eqnarray}
        E &=& -\map A r \,\ell^t \label{eq:kil0E}  \\
        L_x &=& -\map C r \paren{\sin{\phi}\,\ell^\theta+\frac{\sin{2\theta}}{2}\cos{\phi}\,\ell^\phi}  \label{eq:kil0Lx}  \\
        L_y &=& \map C r \paren{\cos{\phi}\,\ell^\theta-\frac{\sin{2\theta}}{2}\sin{\phi}\,\ell^\phi}  \label{eq:kil0Ly}  \\
        L_z &=& \map C r \sin^2{\theta}\,\ell^\phi,  \label{eq:kil0Lz}  
\end{eqnarray}
$\ell^\mu$ denoting the components of the tangent vector.
By convention, we consider here null geodesics parametrized backwards in time, from the observer towards the emitter. For this reason we demand $\ell^t = \frac {dt} {d\lambda} <0$. For simplicity we also fix the parametrization so that the observation point corresponds to $\lambda_\calO = 0$. 

Since $\ell^\mu$ is tangent to a geodesic, its length $\epsilon = g_{\mu\nu}\,\ell^\mu\,\ell^\nu$ is conserved as well:
\begin{equation}
\label{eq:ged}
    \epsilon = - \map A r (\ell^t)^2 + \map B r (\ell^r)^2 + \map C r \paren{\paren{\ell^\theta}^2 + \sin^2 \theta \paren{\ell^\phi}^2} 
\end{equation}
Obviously, photon worldlines correspond to $\epsilon=0$, but we keep $\epsilon$ here unspecified to allow for unconstrained variations of the initial data of the geodesic. Upon the substitution of \eqref{eq:kil0E}-\eqref{eq:kil0Lz} into \eqref{eq:ged} one gets
\begin{equation}
   \epsilon= - \dfrac{E^2}{A} + B\cdot \paren{\ell^r}^2 + \dfrac{L_x^2 + L_y^2 + L_z^2}{C} ,
\label{eq:norm} 
\end{equation}
where $L^2_x + L^2_y + L^2_z = L^2$.
This can be solved for the radial component $\ell^r$:
\begin{equation}
\ell^r = \pm_r {\sqrt{\dfrac{\epsilon}{B} + \dfrac{E^2}{AB} - \dfrac{L^2}{BC}}}. \label{eq:ellr}
\end{equation}
$\pm_r$ here denotes the two possible sign choices for the radial component. This expression for $\ell^r$ in terms of the conserved quantities and $r$ (implicitly present in the metric components $A(r)$, $B(r)$, $C(r)$) will be important later. Since $\ell^r(\lambda) = \frac{d r(\lambda)}{d\lambda}$, \eqref{eq:ellr} can be seen as a first order ODE  for $r(\lambda)$,
which in turn can be solved as an integral with respect to $r$:
\begin{equation}
    \lambda = \fint_{r_\calO}^{r_\calE} \dfrac{\pm_r d \tilde r}{ \sqrt{\dfrac{\epsilon}{B} + \dfrac{E^2}{AB} - \dfrac{L^2}{BC}}}. \label{eq:implicitr}
\end{equation}

We can also solve the equations (\ref{eq:kil0E})-(\ref{eq:ged}) for  $\ell^t$ and integrate the resulting ODE obtaining
\begin{equation}
    t_\calE - t_\calO = -\int_{\lambda_\calO}^{\lambda_\calE} \dfrac{E}{A} d \tilde \lambda = -\fint_{r_\calO}^{r_\calE} \dfrac{E}{A} \dfrac{d \tilde r}{\ell^r}. \label{eq:implicitt}
\end{equation}
We have changed the integration variable to $r$ in the second expression. Recall that in \eqref{eq:ellr} we have expressed $\ell^r$ in terms of $r$ and conserved quantities so that the second integral can be 
evaluated outright.

Now we will consider angular coordinates. From \eqref{eq:kil0E}-\eqref{eq:kil0Lz} we observe that
\begin{equation*}
    \dfrac{d \theta}{\sin^2 \theta} = \paren {\dfrac{L_y}{L_z} \cos \phi - \dfrac{L_x}{L_z} \sin \phi} d \phi,
\end{equation*}
which can be integrated to
\begin{equation}
    \label{eq:gth}
    \cot \theta_\calO - \cot \theta_\calE = \dfrac{L_x}{L_z} \paren {\cos \phi_\calE - \cos \phi_\calO} + \dfrac{L_y}{L_z} \paren{\sin \phi_\calE - \sin \phi_\calO}.
\end{equation}
This implies  that the value of $\theta$ along the geodesic is completely determined by the value of the coordinate $\phi$ and the constants $L_x$, $L_y$ and $L_z$. This is an expression of the fact that the geodesic is contained in a plane orthogonal to $\vec L$. In the final step we solve \eqref{eq:kil0E}-\eqref{eq:ged} for $\ell^\phi$, integrate the resulting ODE and this way derive an implicit expression for $\phi_\calE$:
\begin{equation}
    \int_{\phi_\calO}^{\phi_\calE} \sin^2 \theta d \phi = \int_{\lambda_\calO}^{\lambda_\calE} \dfrac{L_z}{C} d \tilde \lambda = \fint_{r_\calO}^{r_\calE} \dfrac{L_z}{C \ell^r} d \tilde r. \label{eq:implicitphi}
\end{equation}

Equations \eqref{eq:implicitr}-\eqref{eq:implicitphi}, together with \eqref{eq:kil0E}-\eqref{eq:ged}, form the general solution of the geodesic equation in an implicit form.

\subsection{Solution of the geodesic equation in aligned coordinates} \label{sec:aligned}

Before going on, we note that we can also make another use of the large symmetry group of the problem. The geodesic motion in static spherically symmetric spacetime is analogous to the central force problem in Newtonian gravity. Thanks to the $\map {SO} 3$ symmetry, one can rotate the coordinate system to contain the geodesic motion in the $\theta = \pi/2$ plane. This corresponds to angular momentum having only $L_z$ as a non-zero component. In such a coordinate system the geodesic equation is solved by
\begin{equation}
\begin{split}
    \label{eq:geal}
    t_\calE - t_\calO &= -\int_{\lambda_\calO}^{\lambda_\calE} \frac{E}{A} d \tilde \lambda\\
    \lambda_\calE - \lambda_\calO &= \fint_{r_\calO}^{r_\calE} \pm_r \sqrt{\frac{A B C}{A C \epsilon + E^2 C - L_z^2 A}} d \tilde r\\
    \phi_\calE - \phi_\calO &= \int_{\lambda_\calO}^{\lambda_\calE} \frac{L_z}{C} d \tilde \lambda\\
    \theta &= \frac{\pi}{2}
\end{split}    
\end{equation}
In the same way, we may impose the condition $t_\calO = 0$ by applying an appropriate time translation. We will call the coordinate system adapted this way to a given geodesic the \emph{aligned coordinate system}.

\subsection{Method of initial data variations}

We will now apply the method sketched in Section \ref{sec:solution}.
We note here that this is a type of ``brute force" approach to the problem:  it is  algorithmic, but at the same time it requires a lot of algebraic manipulations, involving the differentiation of conservation laws,  solving of systems of linear equations, and matrix multiplication. It may therefore turn out to be unpractical for more complicated metrics without the help of computer-assisted algebra. It is, however, applicable to any metric with a sufficient number of conservation laws.

As we have noted above, the implicit relations defining geodesics consist of two types of equations. In the first group, i.e. \eqref{eq:kil0E}-\eqref{eq:ged}, we have the definitions of five conserved quantities expressed in terms of the initial data $x_\calO^\mu$, $\ell_\calO^\mu$.
We can write them symbolically as
\begin{equation}
J_i = {f_i} (x_\calO^\mu, \ell_\calO^\mu), \label{eq:kilJi}
\end{equation}
with $i=1,\dots,5$. The other group of equations relates the photon's position at $\lambda = \lambda_\calE$ with the initial position and the conserved quantities. These four equations have the form of implicit relations between the coordinates of the point $x_\calE^\mu$ on the geodesic, the corresponding value of the affine parameter $\lambda_\calE$, the initial point $x_\calO^\mu$ and
the conserved quantities $J_i$. The relations 
are implicit, and three of them comprise integrals. In a symbolic form we may write them down as
\begin{equation}\label{eq:hhh}
\begin{split}
h_t(t_{\calE},x_\calO^\mu, J_i,\lambda_\calE) &= 0 \\ 
h_r(r_\calE, x_\calO^\mu, J_i, \lambda_\calE) &= 0 \\
h_\theta(\theta_{\calE}, x_\calO^\mu, J_i, \lambda_\calE) &= 0 \\
h_\phi(\phi_\calE, x_\calO^\mu, J_i, \lambda_\calE) &= 0 
\end{split}
\end{equation}
While the total number of conserved quantities is five, we need to note that the values of the three components of the angular momentum are not entirely independent. Namely, given the vector $\vec y_\calO = (r_\calO\,\sin\theta_\calO\,\cos\phi_\calO,r_\calO\,\sin\theta_\calO\,\sin\phi_\calO,r_\calO\,\cos\theta_\calO)$, defining the photon position in quasi-Cartesian coordinates, we have
\bea
\vec y_\calO(x^\mu_\calO) \cdot \vec L = 0. \label{eq:planarorbit}
\eea
Moreover, this relation, with the same $\vec L$, must hold at all times along a geodesic. Equation \eqref{eq:planarorbit} expresses the fact that all orbits in a spherically symmetric spacetime are planar, with the orbital plane perpendicular to $\vec L$. Now, given the initial point $x^\mu_\calO$, we may use \eqref{eq:planarorbit} to eliminate one of the components of $\vec L$ in favour of the other two. Therefore the total number of independent relations we have obtained is just 8.

 We know from the previous section that we can choose our coordinate system to be aligned, meaning that $\theta_\calO = \frac \pi 2$, $\dot \theta_\calO = 0$, $\phi_\calO = 0$,  
  $t_\calO = 0$ for the fiducial null geodesic $\gamma_0$. We will impose this coordinate condition, but only after the variations are performed, to keep the geodesics' variations unconstrained.
 
 The derivation of $\cal W$ in the coordinate tetrads goes through a sequence of algebraic manipulations of the linear relations between
 the variations of the initial data, the data at $\calE$ and the conserved quantities. 
 These linear relations, in turn, are obtained by taking total variations of the implicit equations above. Note that since all algebraic operations involved in this procedure, i.e.
 substituting and solving for particular variations, are performed at the level of linearized relations, the method always works, although it is, in the end, rather cumbersome.
 
The derivation of $\calW$ proceeds now as follows: we begin by varying the first set of equations, i.e. \eqref{eq:kil0E}-\eqref{eq:kil0Lz}, or symbolically \eqref{eq:kilJi}, obtaining the relations
\bea
\delta J_i = {\cal L}(\delta x_\calO^\mu,\delta \ell_\calO^\nu), \label{eq:deltaJi}
\eea
with $\cal L$ denoting from now on any unspecified linear relation. 
We then vary the second set, i.e. \eqref{eq:implicitr}-\eqref{eq:implicitphi}, or \eqref{eq:hhh}, obtaining after simple manipulations relations of the type
\bea
\delta x_\calE^\mu = {\cal L}(\delta x_\calO^\alpha, \delta J_i, \delta \lambda). \label{eq:deltaxE}
\eea
 After the variation we impose the condition for aligned coordinates in the sense of Section \ref{sec:aligned} for the fiducial null geodesic. This way we simplify the algebraic expressions for the 
coefficients present in both linear relations. We then substitute \eqref{eq:deltaJi} into
 \eqref{eq:deltaxE}, obtaining direct relations between the variations of the initial data and the variations of the final position:
\bea
\delta x_\calE^\mu = {\cal L}(\delta x_\calO^\alpha, \delta \ell_\calO^\beta, \delta \lambda). \label{eq:deltaxEright}
\eea 
We have derived this way the first half of the linear relations we need.

The second half is the linear relations between the variations of $\ell_\calE^\mu$ and the variations of the initial data. We can obtain them from the conservation of $J_i$. Note that the variations $\delta J_i$ are related 
to the variations of the data $(x_\calE^\mu, \ell_\calE^\mu)$ at $\calE$ by the same functional relations as those at the initial point, i.e. we have
\bea
\delta J_i = {\cal L}(\delta x_\calE^\mu,\delta \ell_\calE^\nu), \label{eq:deltaJiviadeltaE}
\eea
with the same coefficients of $\cal L$ as in (\ref{eq:deltaJi}), but evaluated at point $\calE$ instead of $\calO$. 
We may now combine \eqref{eq:deltaJiviadeltaE} with \eqref{eq:deltaJi} and solve the resulting linear equations for $\delta \ell_\calE^\mu$. This yields a relation of type $\delta \ell_\calE^\mu = {\cal L}(\delta x_\calO^\alpha, \delta \ell_\calO^\beta, \delta x_\calE^\gamma, \delta \lambda)$. We now need to eliminate $\delta x_\calE^\gamma$ from this relation using \eqref{eq:deltaxEright} to  obtain
\bea
\delta \ell_\calE^\mu = {\cal L}(\delta x_\calO^\alpha, \delta \ell_\calO^\beta, \delta \lambda). \label{eq:deltalEright}
\eea

By comparing with \eqref{eq:variations}, we note that the coefficients of the linear relations in \eqref{eq:deltaxEright} and \eqref{eq:deltalEright} must be equal to the partial derivatives of the functions $x^\mu(x_\calO^\alpha, \ell_\calO^\beta, \lambda)$ and $\ell^\mu(x_\calO^\alpha, \ell_\calO^\beta, \lambda)$. 
Therefore, in the final step we can use \eqref{eq:covW} to calculate the components of $\calW$ in the coordinate tetrads directly from the coefficients of $\cal L$ in \eqref{eq:deltaxEright} and \eqref{eq:deltalEright}. The result is quite complicated, and we present it in  Appendix \ref{appendix:BGOc}, while the intermediate steps of the calculations are contained in Appendix \ref{app:implicitvar}.  However, as we will see in Section \ref{sec:projections}, it can be simplified by a lot with an appropriate choice of the two tetrads.

\subsection{Geodesic bilocal operators from the solution of GDE by using Killing vectors}

The previous method is very straightforward and could in principle be implemented as an algorithm with any computer algebra program. On the other hand, its manual implementation is extremely tedious. For this reason we present a simpler method which uses directly the GDE and its conserved quantities.

\subsubsection{Overview of the method}

The method of Killing conservation uses the fact that each Killing vector generates the first integral of GDE. The conserved quantites can then reduce the order of the GDE system and the new first order system of ODE's is much easier to solve. In the process, one must introduce integration constants, which can be related to the perturbations of initial data via the GDE conservation equations. Then the components of $W$ operators can be read off one by one.
Unlike the method of initial data variations, where variation had to be done in arbitrary coordinates to retain all the effects of deviation, and only in the very end a particular coordinate choice was set, here we may work in  the aligned coordinates from the very beginning. GDE already contains all effects we are interested in, and we can start in the aligned coordinates without loss of generality.

\subsubsection{Conservation equations}

From \eqref{eq:gdek} we know that Killing vectors generate the first integrals of GDE.
However, equation \eqref{eq:gdek} requires not only the Killing vector, but also its covariant derivatives along the geodesic. We begin with the evaluation of derivatives in the aligned coordinates:
\begin{equation}
\begin{split}
T^\mu &= \paren {1, 0, 0, 0}\\[1ex]
\Phi^\mu_z &= \paren {0, 0, 0, 1}\\[1ex]
\Phi^\mu_x &= \paren{0, 0, -\sin \phi, 0}\\[1ex]
\Phi^\mu_y &= \paren{0, 0, \cos \phi, 0}
\end{split}
\quad
\begin{split}
  \nabla_\ell T^\mu &= \frac{A'}{2} \paren{\frac{\ell^r}{A}, \frac{\ell^t}{B}, 0, 0}\\
 \nabla_\ell \Phi^\mu_z &= \frac{C'}{2} \paren{0, -\frac{~\ell^\phi}{B}, 0, \frac{\ell^r}{C}} \\
\nabla_\ell \Phi^\mu_x &= \paren{0, 0, -\cos \phi \, \ell^\phi - \frac{C'}{2C} \ell^r \sin \phi, 0} \\
\nabla_\ell \Phi^\mu_y &= \paren{0, 0, -\sin \phi \, \ell^\phi + \frac{C'}{2C} \ell^r \cos \phi, 0}
\end{split}
\end{equation}

The first integrals of the GDE in the form of \eqref{eq:gde}, generated by Killing vectors have the following form:
\begin{equation}
\label{eq:gdecce}
\begin{split}
    \Sigma_x &= C \sin \phi \dfrac{d \xi^\theta}{d \lambda} - C \xi^\theta \cos \phi \, \ell^\phi\\
    \Sigma_y &= -C \cos \phi \dfrac{d \xi^\theta}{d \lambda} - C \xi^\theta \sin \phi \, \ell^\phi\\
    \Sigma_z &= -C \dfrac{d \xi^\phi}{d \lambda} - C' \xi^r \ell^\phi\\
    \Sigma_T &= A \dfrac{d \xi^t}{d \lambda} + A' \xi^r \ell^t
\end{split}
\end{equation}

By evaluating \eqref{eq:gdek} at the initial point, where $\xi^\mu = \delta x_{\cal O}^\mu$ and $\nabla_l \xi^\mu = \Delta l_{\cal O}^\mu$, we find expressions of conserved quantities in terms of initial data:
\begin{equation}
\label{eq:sigin}
    \begin{split}
        \Sigma_T &= -\frac{A_\calO'}{2}\ell^r \delta x^t_\calO + \frac{A_\calO'}{2}\ell^t \delta x^r_\calO + A_\calO \Delta \ell^t_\calO\\
        \Sigma_x &= -\paren{C_\calO \cos \phi_\calO \ell^\phi_\calO + \frac{C_\calO'}{2} \ell^r \sin \phi_\calO} \delta x_\calO^\theta + C_\calO \sin \phi_\calO \Delta \ell_\calO^\theta\\
        \Sigma_y &=  \paren {-C_\calO \sin \phi_\calO \ell^\phi_\calO +\frac {C_\calO'} 2 \ell^r \cos \phi_\calO} \delta x_\calO^\theta - C_\calO \cos \phi_\calO \Delta \ell_\calO^\theta\\
        \Sigma_z &= -\frac{C_\calO'}{2} \ell^\phi_\calO \delta x^r_\calO + \frac{C_\calO'}{2} \ell^r_\calO \delta x^\phi_\calO - C_\calO \Delta \ell^\phi_\calO
    \end{split}
\end{equation}
We keep $\phi_\calO$ arbitrary, because setting it to zero at this stage complicates the evaluation of $W$ operators. We leave this value unspecified until $\xi$ is fully expressed in terms of initial data.

Note that the first two equations in \eqref{eq:gdecce} are related, i.e. by shifting $\phi \to \phi - \frac \pi 2$ one can obtain the second equation from the first one. Hence, in order to have $4$ independent first integrals we need include \eqref{eq:inner}, or, to be more exact,  its covariant derivative along $\ell$. Then we have one more first integral:
\begin{equation}
\label{eq:calb}
\begin{split}
    \mathcal B &= E \dot \xi^t + \dfrac{d}{d \lambda} \paren{B \ell^r \xi^r} + L_z \dot \xi^\phi \\
    \mathcal B &= E \Delta \ell^t_\calO + B_\calO \ell^r_\calO \Delta \ell^r_\calO + L_z \Delta \ell^\phi_\calO
\end{split}
\end{equation}
Now we have a sufficient number of equations for integration.

\subsubsection{Solving the equations}
We begin by solving for $\xi^\theta$. Although it looks like we need an explicit form of $\map \phi \lambda$, actually we can integrate it without referring to any particular solution. From \eqref{eq:gdecce} and \eqref{eq:sigin} we have that:
\begin{equation}
 \label{eq:thetasol}
 \begin{split}
    \xi^\theta &= \kappa_1 \sin \phi - \frac{\Sigma_x}{L_z} \cos \phi\\
    \nabla_\ell \xi^\theta &= \xi^\theta \paren{\cot \phi \, \ell^\phi + \frac{C'}{2C} \ell^r} + \frac{\Sigma_x}{C \sin \phi}
    \end{split}
\end{equation}
Here $\kappa_1$ is an arbitrary constant of integration. We see that $\xi^\theta$ depends on $\lambda$ only through $\map \phi \lambda$. This is simply a reiteration of the fact, that in static spherically symmetric spacetimes the dynamics of $\theta$ is constrained by $\phi$. In the same way, the dynamics of the perturbation of $\theta$ is also constrained by $\phi$. 

The other three components are coupled through $\xi^r$. From \eqref{eq:gdecce} and $\eqref{eq:calb}$ we write an equation for $\xi^r$:
\begin{equation}
    \dot \xi^r - \frac{\dot \ell^r}{\ell^r} \xi^r + \frac{1}{B \ell^r} \paren{\frac{E \Sigma_T}{A} - \frac{L_z \Sigma_z}{C} - \mathcal B} = 0
\end{equation}
Integrating it yields a solution for $\xi^r$, which is then used to find $\xi^t$ and $\xi^\phi$:
\begin{equation}
\begin{split}
        \xi^r &= \kappa_2 \ell^r - \ell^r \int_{\lambda_\mathcal O}^{\lambda_\mathcal E} \paren{\frac{E \Sigma_T}{A} - \frac{L_z \Sigma_z}{C} - \mathcal B} \dfrac{d \lambda}{B {\ell^r}^2}\\
     \xi^t &= \kappa_3 + \int_{\lambda_\calO}^{\lambda_\calE} \paren{\frac{\Sigma_T}{A} + \frac{E A'}{A^2} \xi^r} d\lambda\\
     \xi^\phi &= \kappa_4 - \int_{\lambda_\calO}^{\lambda_\calE} \paren{\frac{\Sigma_z}{C} + \frac{C'}{C^2} L_z \xi^r} d \lambda
    \end{split}
    \end{equation}
where $\kappa_2, \kappa_3, \kappa_4$ are arbitrary constants. The last step is to express all the constants in terms of initial data -- $\delta x^\mu_\calO$ and $\Delta \ell^\mu_\calO$. This can be done by evaluating $\xi^\mu$ and $\nabla_\ell \xi^\mu$ at $\calO$ together with \eqref{eq:sigin} and \eqref{eq:calb}. Then the $W$ operators can be found by comparing $\xi^\mu$ and $\nabla_\ell \xi^\mu$ with Eqs. \eqref{eq:var01} and \eqref{eq:var02}. Formally, we can write
\begin{equation}
    {W_{XX}} \UD \mu \nu = \frac{\partial \xi^\mu}{\partial \delta x^\nu_\calO} \quad {W_{XL}} \UD \mu \nu = \frac{\partial \xi^\mu}{\partial \Delta \ell^\nu_\calO} \quad {W_{LX}} \UD \mu \nu = \frac{\partial \nabla_\ell \xi^\mu}{\partial \delta x^\nu_\calO} \quad {W_{LL}} \UD \mu \nu = \frac{\partial \nabla_\ell \xi^\mu}{\partial \Delta \ell^\nu_\calO}.
\end{equation}
Explicit expressions of all these components with respect to the aligned coordinate tetrad can be found in the Appendix.

\subsection{Construction of a parallel propagated tetrad} \label{sec:parallel}

In the previous sections we described how to obtain an exact solution to the GDE for static spherically symmetric spacetimes in the coordinate tetrad. However, due to the diffeomorphism invariance of General Relativity, physical aspects of geodesic deviation are obscured by the choice of coordinates. In order to mitigate this problem we will project our results onto a  parallel propagated tetrad. In this paper we will use the SNT \cite{grasso2019}, while the construction itself is based on \textcite{marck1983}. For more details on the complete integrability of parallel transport please check the review in \cite{frolov2017}.

The \emph{semi-null tetrad} (SNT) $e \UD \mu {\boldsymbol \mu} = \paren {u^\mu, e^\mu_\mba, \ell^\mu}$ comprises the four-velocity $u^\mu$, a null vector $\ell^\mu$ and two mutually orthogonal spacelike vectors $e_\mba^\mu$, called the transverse vectors, which are also orthogonal to $\ell^\mu$ and $u^\mu$. This frame is defined by the following constraints:

\begin{equation}
\label{eq:cons}
    \begin{split}
        & \ell^\mu \ell_\mu = 0 \\
        & e^\mu_\mba \ell_\mu = 0 \\
        & e^\mu_\mba e_{\mbb \mu} = \delta_{\mba \mbb}\\
        & u^\mu_\calO u_{\calO \mu} = -1 \\
        & u^\mu_{\calO} e_{\mba \mu} = 0 \\
        & \ell^\mu u_{\calO \mu} = Q > 0
    \end{split}
\end{equation}
where $Q$ is a constant related to the normalization of the null tangent $\ell^\mu$. The construction of the frame will be done in two steps. Firstly, we will reduce the space of tetrads to a subspace of those whose two vectors are parallel propagated. This will leave a one-parameter family of tetrads at each point. Then we will use a linear transformation together with the parallel propagation equation to parallel transport the entire tetrad.

We begin the first step with an observation that $\ell^\mu$ is already parallel propagated. To obtain the second vector, we notice that the vector that is perpendicular to the plane passing through the origin of the coordinate system is parallel propagated along the whole plane. In our case, this vector is $e^\mu_{\mathbf 1} = \frac 1 {\sqrt C} \partial_\theta$ which is spacelike and perpendicular to the plane $z = 0$.

Now we will seek the vector $e^\mu_{\mathbf 2}$. By looking at Eqs. (\ref{eq:cons}) we see that from conditions $e_{\mathbf 1} \cdot e_{\mathbf 2} = 0$, $e_{\mathbf 2} \cdot \ell = 0$ and $e_{\mathbf 2} \cdot e_{\mathbf 2} = 1$ we get $e_{\mathbf 2}$ up to an additive term $\map c \lambda \cdot \ell$. Hence, from purely geometric considerations and without solving \emph{any} ODEs we get an equivalence class of vectors $e_{\mathbf 2}$ such that for all of them $\nabla_\ell e_{\mathbf 2}$ differs only by $\map f \lambda \cdot \ell$ for some $f$. Next, we pick a particular instance of $e_{\mathbf 2}$, say, $\tilde e_{\mathbf 2} = \map \alpha \lambda \partial_t + \map \beta \lambda \partial_r$, which is not necessarily parallel transported. Then, up to an overall sign, $\tilde e_{\mathbf 2}$ reads
\bea
\tilde e_{\mathbf 2}^\mu = \paren {\sqrt{\frac{BC}{A}}\frac{\ell^r}{L_z}, -\sqrt{\frac{C}{AB}}\frac{E}{L_z},0,0}
\eea
while the unique associated four-velocity $\tilde u^\mu$, orthogonal to both transverse vectors, is of the form
\bea
\tilde u^\mu = \paren {\frac{E}{2AL_z^2Q} \paren{L^2 + CQ^2}, - \frac{\ell^r}{2L_z^2 Q},0,\frac{Q}{2L_z} - \frac{L}{2CQ}}.
\eea
We will call the tetrad $(\tilde u^\mu,e_{\bm 1}^\mu,\tilde e_{\bm 2}^\mu,\ell^\mu)$ the \emph{intermediate} SNT.

In the second step we will look for a linear $\lambda$-dependent transformation that preserves Eqs. (\ref{eq:cons}). Let $\paren {\tilde u^\mu, e^\mu_{\mathbf 1}, \tilde e^\mu_{\mathbf 2},\ell^\mu}$ and $\paren {u^\mu, e^\mu_{\mathbf 1}, e^\mu_{\mathbf 2},\ell^\mu}$ be SNTs, with the second one being parallel transported. We assume the following ansatz:

\begin{equation}
  \paren{\begin{array}{c}
  u^\mu \\
  e_{\mathbf 1}^\mu \\
  e_{\mathbf 2}^\mu \\
  \ell^\mu
  \end{array}}
  = \paren{  \begin{array}{cccc}
         a_u & b_u & c_u & d_u \\
        a_1 & b_1 & c_1 & d_1 \\
        a_2 & b_2 & c_2 & d_2 \\
        a_\ell & b_\ell & c_\ell & d_\ell
    \end{array}} \paren{\begin{array}{c}
          \tilde{u}^\mu \\
  e_{\mathbf 1}^\mu \\
  \tilde{e}_{\mathbf 2}^\mu \\
  \ell^\mu
  \end{array}}
\end{equation}
Since $\ell^\mu$ and $e^\mu_{\mathbf 1}$ are already parallel propagated, some coefficients can be set to either zero or one. The rest of the coefficients are determined by the SNT constraints and can be shown to depend on only one function $\map \Psi \lambda$. Then, up to a sign, $e_{\mathbf 2}^\mu$ is
\bea
e_{\mathbf 2}^\mu = \tilde e_{\mathbf 2}^\mu - \frac{\Psi}{Q}\ell^\mu
\eea
while $u^\mu$ is unique and of the form
\bea
u^\mu = \tilde u^\mu + \Psi \tilde e_{\mathbf 2}^\mu - \frac{\Psi^2}{2Q}\ell^\mu.
\eea
To determine $\Psi$, we have to use the parallel propagation equation for any single vector of the SNT. For example, demanding $\nabla_\ell e_{\mathbf 2}^\mu = 0$ yields
\bea
\dot \Psi \frac{\ell^\mu}{Q} = \nabla_\ell \tilde e_{\mathbf 2}^\mu
\eea
where dot denotes the derivative with respect to $\lambda$. This has to hold for any component. For example, the equation for component $\phi$ yields the following simple ODE:
\bea
\dot \Psi = -\frac{C'}{2\sqrt{ABC}}\frac{E}{L_z}Q
\eea
The initial condition for $\Psi$ will now fix a particular choice of the solution. The natural choice is  $\tilde e_{\mathbf 2}$ at the initial point, leading to $\map \Psi 0 = 0$, and this is what we use in the next section, but other choices are possible too.

\subsection{Projections of operators onto a semi-null tetrad} \label{sec:projections}

In the SNT all four BGO's have the following form:

\begin{equation}
\begingroup
   \renewcommand*{\arraystretch}{1}
   {W_{**}} \UDb \mu \nu = \paren{\begin{array}{c|cc|c}
        \alpha & 0 & 0 & 0 \\ \hline
        0 & \blacksquare & 0 & 0 \\
        \square & 0 & \blacksquare & 0 \\ \hline
        \square & 0 & \square & \alpha
   \end{array}}
   \endgroup
\end{equation}
Here $\alpha$ denotes $\lambda$, $1$, $1$ and $0$  for the operators $\WXL$, $\WXX$, $\WLL$, $\WLX$  respectively. Due to Eqs. \eqref{eq:additionalrelations}, the top row and the rightmost column have a fixed form irrespective
of the spacetime geometry. The $2 \times 2$ submatrix in the center corresponds to the projection of the BGOs onto the screen space. It is diagonal in the SNT we have constructed, and all of its components are independent of the choice of the observer's four-velocity $u$ as the
first component of the tetrad. For this reason, the two nonvanishing components, denoted by $\blacksquare$, do not contain $Q$ or $\Psi$. 
This does not apply to the other three nonzero components, denoted by $\square$. Altogether, there are at most five nontrivial components per BGO, but sympletic properties impose 7 constraints, which brings the total number of nontrivial independent components down to 13. We present all the nontrivial components of the BGOs and the optical tidal matrix in the aligned coordinate tetrad and the SNT in the Appendix \ref{appendix:BGOs}. 

A portion of our results has been derived earlier, but in different contexts. In \cite{dwivedi1972} the authors studied the luminosity of a spherically collapsing star. By following a bundle of light coming from a surface area element of the star which reached the observer and integrating over the whole surface of the star, they managed to show the dependence of the total observed flux on the radius of the surface. In \cite{dyer1977} the topic of the study was the spherical gravitational lens and its modification due to clumpiness of the matter within the lens or a large scale matter distribution surrounding the lens itself.

\section{Optical distance measures and distance slip in Schwarzschild spacetime}

Most methods of distance determination in astronomy we know use light propagation one way or another. In a flat spacetime, with no relative motions of the sources and  all distance measures are perfectly equivalent. In general, however, light propagation is affected by the spacetime curvature, which makes distance measures differ from each other and their counterparts in flat spacetime. This leads to many paradoxical results, such as the finite maximal value of the angular diameter distance in many FLRW Universe models \cite{mccrea1935, ellis1985, araujo2009} or absence of parallax in some others \cite{hasse1988}.
In this section, we will to present and analyze in detail the behaviour of the distance measures in static spherically symmetric spacetimes with the help of the results of the previous sections. Although this class of metrics is quite special, it is also sufficiently broad and can be used to study the nontrivial behaviour of distance measures. Even though the final results are complicated, everything can be explained exactly.
 
The two distance measures we consider in this paper are the \emph{angular diameter distance} (also known as the \emph{area distance} \cite{perlick2004r}) $D_{ang}$ and the \emph{parallax distance}  $D_{par}$ \cite{grasso2019}. Just for completeness we recall also the notion of the \emph{luminosity distance} $D_{lum}$ \cite{etherington1933, etherington2007, perlick2004r}, defined with the help of the measured flux of energy from a radiating body of known luminosity. It is well-known that it is related to $D_{ang}$ via the Etherington's duality relation \cite{perlick2004r}.
$D_{ang}$ on the other hand is defined using the ratio of the solid angle an extended object takes up on the observer's celestial sphere to its physical cross-sectional area. The definition of $D_{par}$ on the other hand makes use of the trigonometric
parallax effect, i.e. the dependence of the position of the source's image on the celestial sphere on the observer's transverse displacement suitably averaged over the baseline orientation. We will also discuss
the distance slip $\mu$, introduced in \cite{grasso2019} and defined as the relative difference between $D_{ang}$ and $D_{par}$. It is an interesting quantity because it directly measures the impact of the spacetime curvature on the light propagation in a frame-independent way. Moreover, for short distances $\mu$ is equal to an integral of the mass density along the LOS.

The key observation is that the distance slip and the two distance measures between two points connected by a null geodesic $\gamma_0$ can be expressed in terms of the BGOs and the observer's four-velocity \cite{grasso2019}.
This means that the BGO formalism can be used to investigate the dependence of the distance measures on the null geodesic $\gamma_0$ and the positions of the emission and observation point along it. 

After introducing the distance measures and the infinitesimally thin ray bundle formalism, we will consider the simplest nontrivial example of static spherically symmetric spacetimes, namely, the Schwarzschild black hole. We will numerically study the behaviour of ray bundles that begin at some distance from the photon sphere, propagate towards and around the black hole, possibly, winding around it a number of times, and, finally, escape to infinity. The form of trajectories will be controlled by the initial data of the fiducial geodesic. Using the results of this paper, we will discuss the properties of $D_{ang}$, $D_{par}$ and $\mu$ associated with sources positioned at different points along the null geodesic. Finally, we will prove a few more general statements regarding to the behaviour of these functions. Since distance measures can be expressed in terms of $W$, we have all the tools to study them in general.

\subsection{Distance measures}

We begin with a short review of the distance measures. We consider here three distinct definitions of distance: the angular diameter distance $D_{ang}$, the parallax distance $D_{par}$ and the luminosity distance $D_{lum}$. In this section we will build upon the definitions provided in \cite{grasso2019}.

In the gravitational lensing theory, $D_{ang}$ is obtained from the magnification matrix $M \UD \mba \mbb$, which relates the physical size of the source $\delta x^\mbb_\calE$ to its angular size at the observer's sky $\delta \theta^\mba_\calO$. In the parallel propagated SNT this reads:
\bea
\delta \theta^{\mathbf A}_\calO = M \UD \mba \mbb \delta x^\mbb_\calE,
\label{eq:mmrel}
\eea
where
\bea
M \UD \mba \mbb = \paren {\ell_\calO \cdot u_\calO}^{-1} \paren{\DD^{-1}} \UD \mba \mbb
\label{eq:mm}
\eea
and $\DD \UD {\mathbf A} {\mathbf B}$ is the Jacobi map. However, $\DD \UD {\mathbf A} {\mathbf B}$ can also be expressed as the transverse part of $\WXL$, i.e.
\bea
\DD \UD \mba \mbb = {\WXL} \UD \mba \mbb.
\label{eq:jm}
\eea
Then $D_{ang}$ is defined via the magnification matrix in the following way:
\bea
D_{ang} = \size {\det M \UD \mba \mbb}^{-1/2} = \paren {\ell_\calO \cdot u_\calO} \size {\det {\WXL} \UD \mba \mbb}^{1/2}.
\label{eq:dang}
\eea

On the other hand, there are many different possible definitions of the parallax distance \cite{sachs1961, kasai1988, weinberg1970, rosquist1988}, the main difference being the scheme of averaging over the baseline orientation. We will consider the one given in \cite{grasso2019}. Namely, we are interested in the parallax distance obtained by a family of comoving observers receiving light from a single spacetime event. Analogously, we introduce the parallax matrix $\Pi \UD \mba \mbb$ that relates the transverse displacement $\delta x^\mbb_\calO$ of a secondary observer with respect to the one at $\calO$ and the angular variation of the image on the observer's sky, $\delta \theta^\mba_\calO$: 
\bea
\delta \theta^{\mathbf A}_\calO = - \Pi \UD \mba \mbb \delta x^\mbb_\calO
\label{eq:par}
\eea 
where
\bea
\Pi \UD \mba \mbb = \paren {\ell_\calO \cdot u_\calO}^{-1} \paren{\DD^{-1}} \UD \mba \mbc \paren {\delta \UD \mbc \mbb + {m_\perp} \UD \mbc \mbb}
\label{eq:parm}
\eea
and ${m_\perp} \UD \mba \mbb$ is the transverse part of the emitter-observer asymmetry operator \cite{grasso2019, korzynski2018}. In the BGO language:
\bea
{\WXX} \UD \mba \mbb = \delta \UD \mba \mbb + {m_\perp} \UD \mba \mbb.
\eea
Then $D_{par}$ is defined via the parallax matrix in the following way:
\bea
D_{par} = \size {\det \Pi \UD \mba \mbb}^{-1/2} = \paren {\ell_\calO \cdot u_\calO} \size {\det {\WXL} \UD \mba \mbb}^{1/2} \size {\det {\WXX} \UD \mba \mbb}^{-1/2}.
\label{eq:dpar}
\eea
Both distance measures depend on the BGOs, and thus on the spacetime curvature along the LOS, and the observer's four-velocity $u_{\calO}^\mu$ (due to the light aberration effect), but not
on the source's four-velocity $u_\calE^\mu$.

Let us also remind the luminosity distance $D_{lum}$, defined operationally by comparing the observed electromagnetic energy flux to the absolute luminosity of the source \cite{perlick2004r}. 
Unlike the previous two, it depends on both the emitter's and observer's four-velocity. It is well-known that it is not independent of the other two:
the Etherington's reciprocity relation states that it can be expressed in terms of  $D_{ang}$ and the redshift measured by the observer:
\bea
D_{lum} = D_{ang} \paren {1 + z}^2.
\eea
For this reason we do not consider it in this paper.

The distance slip is defined via
\bea
\mu = 1 - \sigma \frac{D_{ang}^2}{D_{par}^2}, \label{eq:mudef1}
\eea
where the sign
\bea
\sigma = \sgn \det {\WXX} \UD \mba \mbb. \label{eq:mudef2}
\eea
Unless we are dealing with strong lensing, the sign $\sigma$ is equal to $+1$. Therefore, in most cases, $\mu$ measures the difference between the results of two distance measurements to a single object.  
It vanishes identically in the Minkowski spacetime or whenever the optical tidal matrix $\eqref{eq:otm}$ vanishes along the LOS. In general, it is expressible as a functional of curvature tensor along the LOS. 
The distance slip has an important property of being independent of any kinematical variables describing momentary motions of both the source and the observer, for example, their four-velocities $u_\calO^\mu$, $u_\calE^\mu$. Just like $D_{ang}$ and $D_{par}$, it has a simple expression in terms of the transverse components of the BGOs:
\bea
\mu = 1 - \det {\WXX} \UD \mba \mbb.
\eea

In the Minkowski spacetime, in the absence of relative motion, $D_{lum}$, $D_{ang}$ and $D_{par}$ are equivalent. In the general case, however, the behaviour of these distances for objects far away can be rather complicated and paradoxical. The results of this paper allow us to discuss it thoroughly and understand it at least in static spherically symmetric spacetimes. In particular, we may consider the physically important case of observing a source near a large mass represented by a Schwarzschild black hole.

\subsection{Infinitesimally thin bundles}

The behaviour of distance measures is much simpler to grasp if we relate it to the properties of infinitesimally thin bundles of light rays. For this reason we briefly review the basics of ray bundle formalism.
We follow the definitions and conventions of \textcite{perlick2004r}.  Let $\lambda \mapsto \map x \lambda$ be an affinely parametrized null geodesic with a tangent vector $\ell = \dot x$. An \emph{infinitesimally thin bundle of rays}  is the set
\bea
S = \set {c^I \xi \UD {\bsmu} I ~|~ c^1,c^2 \in \RR, \, c^I c^J \delta_{IJ} \le 1}
\eea
where, in the parallel propagated SNT, $\xi \UD \bsmu I$ satisfies the GDE (cf. \eqref{eq:gde})
\bea
\ddot \xi \UD {\boldsymbol \mu} I &= \calR \UD {\boldsymbol \mu} {\boldsymbol \nu} \xi \UDD {\boldsymbol \nu} I, 
\label{eq:gdee}
\eea
together with the orthogonality constraint
\bea
g_{\bsmu \bsnu} \ell^{\bsmu} \xi \UD {\bsnu} I = 0 \label{eq:ort}
\eea
and $I$ enumerates linearly independent solutions.
By construction, its cross-section by the screen space of an observer is elliptical and spacelike. This problem setting is equivalent to the one used to prove the Sachs shadow theorem \cite{sachs1961}, where a small object in a null geodesic congruence casts a shadow on a screen in motion. In either case, the area of this cross-section is a Lorentz invariant at any given point of the geodesic, i.e. it does not depend on the observer we choose. 
The area can be expressed as
\bea 
\calA = \int_\Sigma \epsilon_{\mba \mbb} \xi^\mba_1 \xi^\mbb_2,
\label{eq:ardef}
\eea 
where $\epsilon_{\mba\mbb}$ is the area two-form,   
and $\xi^\mba_I$ are the projections of linearly independent solutions of \eqref{eq:gdee} onto the screen space, i.e. space spanned by the two transverse vectors in a SNT. It evolves according to the equation
\bea
\dfrac {d \calA} {d \lambda} = \calA\, \theta,
\label{eq:area}
\eea
where $\theta$ is the bundle expansion. Note that the area defined this way is a signed quantity. The sign can change every time the bundle degenerates to a line or a point. 

In order to determine $\theta$ one has to make use of null Raychaudhuri equations (also known as Sachs optical equations). In this paper we will only consider the twist-free (or surface-forming) bundles, i.e. those for which the twist $\omega_{\mba\mbb}$ vanishes. The equations read \cite{poisson2004}
\bea
\dfrac{d \theta}{d \lambda} = - \frac{\theta^2}{2} - \sigma_{\mba \mbb} \sigma^{\mba \mbb} - R_{\mu\nu} \ell^\mu \ell^\nu
\label{eq:raye}
\eea
\bea
\dfrac{d \sigma_{\mba \mbb}}{d \lambda} = - \theta \,\sigma_{\mba \mbb} + C_{\mba \mu \nu \mbb} \ell^\mu \ell^\nu,
\label{eq:rays}
\eea
which in the BGO formalism are equivalent to
\bea
\dfrac{d^2}{d \lambda^2} {W_{**}} \UD \mba \mbb = \paren {- \frac 1 2 R_{\mu \nu}\ell^\mu \ell^\nu \delta \UD \mba \mbc + C \UDDD \mba \mu \nu \mbc \ell^\mu \ell^\nu} {W_{**}} \UD \mbc \mbb.
\label{eq:gdepp} 
\eea
For the purpose of this paper we introduce two infinitesimal ray bundles along $\gamma_0$: the \emph{vertex bundle} and the \emph{initially parallel bundle}. The vertex bundle is a bundle of rays crossing at $\calO$. It is defined by the singular initial conditions for $\theta$, $\sigma_{\mba\mbb}$ and $\omega_{\mba\mbb}$ at $\calO$ \cite{perlick2004r}:
\bea
 \theta(\lambda) &\sim& \frac 2 {\lambda - \lambda_{\calO}}\\
 \sigma_{\mba\mbb}(\lambda_\calO) &=& 0\\
 \omega_{\mba\mbb}(\lambda_\calO) &=& 0.
\eea
The initially parallel bundle, on the other hand, is strictly parallel at $\calO$, i.e.
\bea
 \theta(\lambda) &=& 0 \\
 \sigma_{\mba\mbb}(\lambda_\calO) &=& 0 \\
 \omega_{\mba\mbb}(\lambda_\calO) &=& 0 .
\eea
Both bundle are twist-free, or surface forming, i.e. $\omega_{\mba\mbb}=0$ along the whole null geodesic. They are both closely related to the transverse components of the operators $\WXX$ and $\WXL$: namely, we have
\bea
\xi^{\mba}(\lambda) = {\WXL}\UD{\mba}{\mbb}(\lambda)\,\dot\xi^{\mbb}(\lambda_\calO) \label{eq:xiWXL}
\eea
for the vertex bundle and 
\bea
\xi^{\mba}(\lambda) = {\WXX}\UD{\mba}{\mbb}(\lambda)\,\xi^{\mbb}(\lambda_\calO) \label{eq:xiWXX}
\eea
for the initially parallel bundle. Note that due to the orthogonality condition (\ref{eq:ort}) $\xi^\mu$ has only transverse components plus a component proportional to $\ell^\mu$. The latter is irrelevant
from the point of view of the geometry of cross-sections (see \cite{sachs1961}), so it is the two transverse components of $\xi^\mu$ given by (\ref{eq:xiWXL})-(\ref{eq:xiWXX}) that define the 
distance measures.

Having developed the ray bundle formalism, now we can utilize it to understand distance measures solely in terms of the cross-sectional areas of various ray bundles. 

Let us begin with the vertex bundle. From \eqref{eq:mmrel}-\eqref{eq:jm} and $\eqref{eq:xiWXL}$ it follows that
\bea
\delta \theta^\mba_\calO = (\ell_\calO \cdot u_\calO)^{-1} \dot \xi^\mba_\calO.
\eea
Integrating \eqref{eq:mmrel} over the angular shape of the figure on the observer's sky yields
\bea
\tilde \calA (\lambda) = \paren {\det M \UD \mba \mbb}^{-1} \tilde \Omega_\calO 
\eea 
where $\tilde \Omega_\calO$ is the angular area of the figure as observed from $\calO$, and $\tilde A (\lambda)$ is the physical area of the cross-section of the ray bundle at $\calE$. From $\eqref{eq:dang}$ it simply follows that
\bea
D_{ang} = \sqrt{\frac{|\tilde \calA (\lambda)|}{\tilde \Omega_\calO}}.
\eea
Previously introduced initial conditions for the vertex bundle imply that its cross-sectional area at $\calO$ exhibits the following behaviour:
\bea
\map {\tilde \calA} \lambda = \paren{\lambda - \lambda_\calO}^2 \paren{\ell_\calO \cdot u_\calO}^2 \tilde \Omega_\calO + \map \calO {\lambda^3}.
\eea
Consider now the initially parallel bundle. Its evolution is described by the $\WXX$ operator, which stands for the following mapping:
\bea
\xi^\mba (\lambda) = {\WXX} \UD \mba \mbb (\lambda) \,\xi^\mbb (\lambda_\calO).
\label{eq:wxx}
\eea
Suppose the cross-sectional area of this bundle at $\calO$ is $\calA_\calO$. Integration of \eqref{eq:wxx} over the initial shape of the cross-section gives \cite{korzynski2021}
\bea
\map \calA \lambda = \paren {\det {\WXX} \UD \mba \mbb} \calA_\calO,
\eea
which allows us to rewrite $\mu$ as
\bea
\mu = 1 - \frac{\map \calA \lambda}{\calA_\calO}
\label{eq:muar}
\eea
with $\map \calA {\lambda_\calO} = \calA_\calO$.
Finally, the substitution of the results presented above into $\eqref{eq:mudef1}$ enables us to write down the parallax distance:
\bea
D_{par} = \sqrt{\size {\frac{\map{\tilde \calA} \lambda}{\map \calA \lambda}}} \sqrt{\frac{\calA_\calO}{\tilde \Omega_\calO}}.
\eea
To sum up, the ray bundle formalism allows us to express distance measures and their slip through cross-sectional areas in a relatively simple way. The analysis can be made even more straightforward if we apply the BGO formalism.

\subsection{Special points}

We fix the null geodesic $\gamma_0$ and the observer's position $\calO$ along $\gamma_0$. We can now introduce three types of special points along a  null geodesic, defined by the properties of the vertex and initially parallel ray bundle. Their importance stems from the fact that they mark points where  $D_{ang}(\lambda)$, $D_{par}(\lambda)$ and $\mu(\lambda)$ take particular values. Each of these points may appear an arbitrary number of times along a null geodesic or not appear at all, depending on the spacetime geometry.

We call $\cal P$ a \emph{conjugate point} with respect to $\calO$ iff the vertex bundle from $\calO$ refocuses back at
$\cal P$ at least along one transverse direction. This property is equivalent to the existence of a Jacobi field along $\gamma_0$, vanishing at $\cal O$ and $\cal P$, but not identically zero. It is easy to check that this happens iff $\det {\WXL}\UD{\bf A}{\bf B} = 0$ between $\calO$ and $\calP$. Conjugate points correspond to the intersection of the fiducial geodesic with a caustic and are points of infinite magnification
of images of objects located at $\cal P$ as seen in $\calO$. We can see that in these points we formally have $D_{ang} = 0$. Moreover, as long as $\det  {\WXX}\UD{\bf A}{\bf B} \neq 0$, we also have $D_{par} = 0$. On the other hand, $\mu$ does not need to take any special value in a conjugate point because its value is unrelated to the properties of the vertex bundle.

We call $\cal P$ a \emph{focal point} iff an infinitesimal bundle of rays running parallel at $\cal O$ refocuses at $\cal P$ along at least in one direction.
This happens when $\det  {\WXX}\UD{\bf A}{\bf B} = 0$. The physical interpretation of these points is vanishing parallax effect along at least one baseline: the parallax matrix must be degenerate in at least one direction. This means that the displacements of the observer in this direction result in no measurable image displacement. It is straightforward to see that at these points we have diverging parallax distance, i.e.  $D_{par} \to \infty$, as long as $\cal P$ is not a conjugate point as well. Moreover, at focal points we always have
$\mu = 1$. However, the value of $D_{ang}$ can be arbitrary at a focal point. 

Finally, $\cal P$ is an \emph{equidistance point}  iff $D_{ang} = D_{par}$, i.e. both methods of distance determination yield the same value. The reader may check easily from \eqref{eq:mudef1}-\eqref{eq:mudef2} that at these points we have either $\mu = 0$ or $\mu = 2$.

\subsection{Numerical results}
Now we will use this formalism to study light propagation and notions of distances in Schwarzschild spacetime. In this analysis we are interested in a beam of light that connects static emitters and observers placed outside the black hole's photon sphere. The trajectory of the geodesic is an arbitrary arc that (possibly) winds around the black hole a finite number of times. We fix the observer's position at $r = 100 r_s$, where $r_s$ is the Schwarzschild radius, and vary the impact parameter $b = \frac{\size {L_z}}{E}$.  Then we follow the corresponding null geodesic as we increase the affine parameter value $\lambda$, and to each value we assign an emitter placed at the point $\map {x^\mu} \lambda$. The parametrization of the geodesic is fixed by rescaling the affine parameter while keeping the products $\paren{u_\calO \cdot \ell_\calO}$ equal to $1$ for all instances of $b$. This means that $\lambda$ agrees with the spatial distance measured by the observer in his or her vicinity. In every case the endpoint of the geodesic is placed sufficiently far away from the black hole such that the curvature effects eventually would be negligible. The plots reveal that the evolution of distance measures and $\mu$ has three distinct stages. For this reason we will discuss their behaviour in the initial, intermediate and faraway regions separately.

\begin{figure}[H]
\begin{subfigure}[b]{.49\linewidth}
\centering
\includegraphics[width=\linewidth]{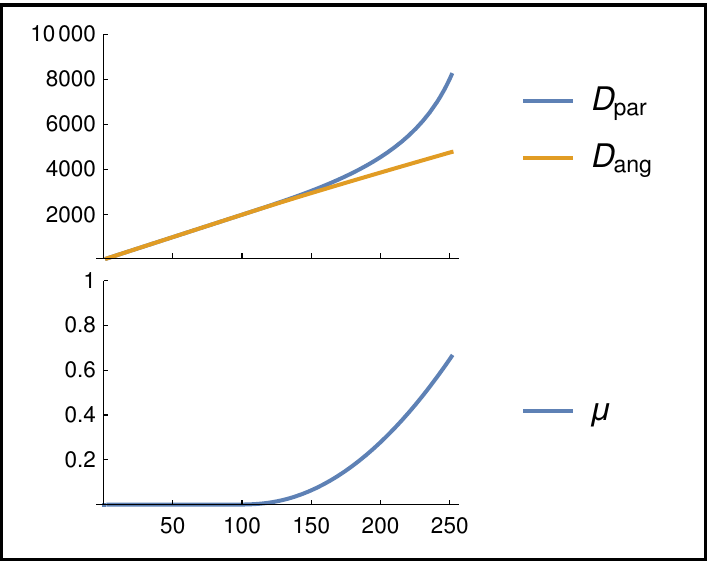}
\caption{$r_\calO = 100 ~r_s$, $b = 20~r_s$}\label{fig:fig1a}
\end{subfigure}\hfill
\begin{subfigure}[b]{.49\linewidth}
\centering
\includegraphics[width=\linewidth]{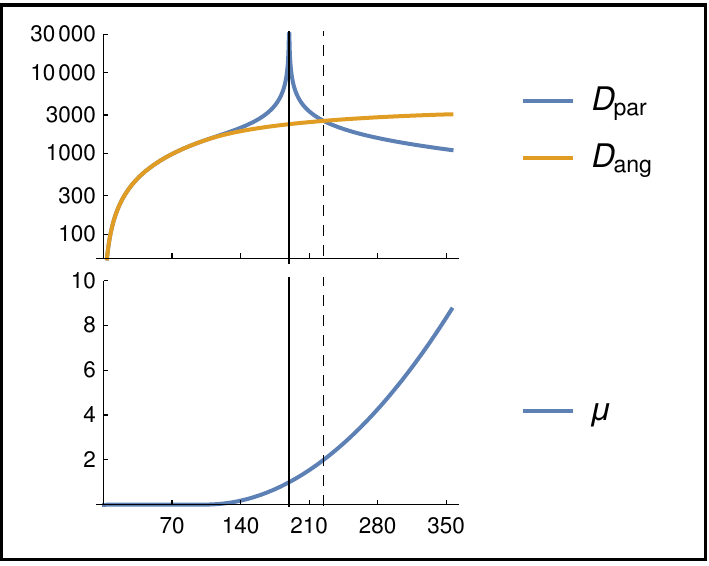}
\caption{$r_\calO = 100 ~r_s$, $b = 14.14~r_s$}\label{fig:fig1b}
\end{subfigure}\hfill
\begin{subfigure}[b]{.49\linewidth}
\centering
\includegraphics[width=\linewidth]{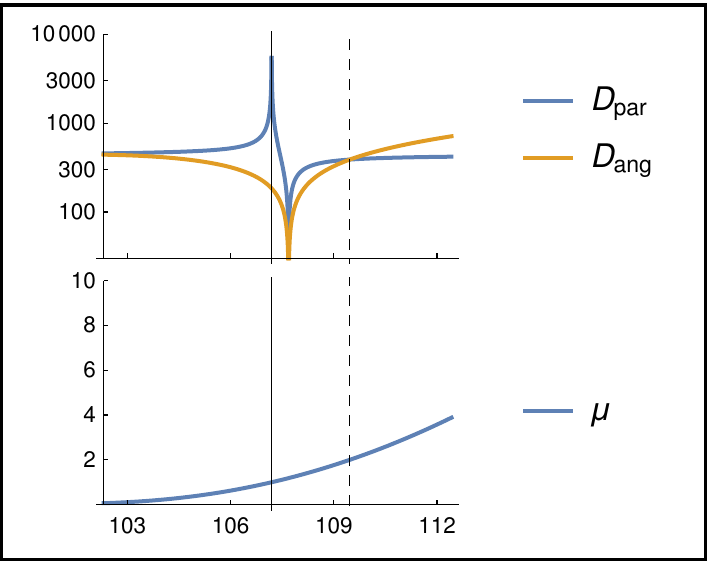}
\caption{$r_\calO = 100 ~r_s$, $b = 4.47~r_s$}\label{fig:fig1c}
\end{subfigure}\hfill
\begin{subfigure}[b]{.49\linewidth}
\centering
\includegraphics[width=\linewidth]{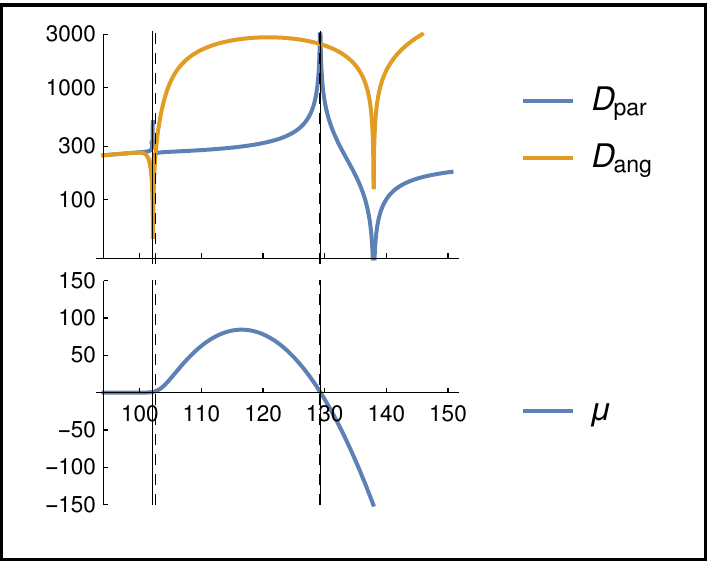}
\caption{$r_\calO = 100 ~r_s$, $b = 2.67~r_s$}\label{fig:fig1d}
\end{subfigure}%
\caption{
Dependence of $D_{par}$, $D_{ang}$ and $\mu$ on initial conditions. The horizontal axis shows the value of the affine parameter along the geodesic. Parametrization is the same for all cases and  is fixed by rescaling the affine parameter in a way that makes the product $\paren{u_\calO \cdot \ell_\calO} = 1$. All distance measures are measured in Schwarzschild radii. Vertical solid lines denote $D_{par} = \infty~(\mu = 1)$, dashed lines -- $D_{par} = D_{ang}$ $\paren{\mu = 1 \pm 1}$. In the last picture the region between the first pair of lines (which almost appears as a single line) shows a behaviour similar to $b = 4.47 r_s$ case. The second group comprises a solid line surrounded by two dashed lines. The distances are plotted in a linear scale in plot (a) and in logarithmic scale in (b), (c) and (d).}
\end{figure}

~

In principle there are two ways to investigate this problem numerically. The first approach is the numerical evaluation of the exact solutions of the geodesic equations and the GDE. The second approach is the direct numerical integration of the geodesic equation and the GDE projected onto the SNT. We choose the second approach because it provides an easier control of the problem, especially at the turning points.

As depicted in Figure \ref{fig:fig1a}, $D_{par}$ and $D_{ang}$ differ very slightly in the case of a large impact parameter. As $\lambda$ approaches $0$, distances become arbitrarily close to each other. On the other hand, the growth of $\lambda$ is accompanied by a slight increase in the difference between both distance measures, with $D_{par}$ being the larger one. Similarly, $\mu$ is slowly monotonically increasing.

A slight decrease of the impact parameter (Figure \ref{fig:fig1b}) results in the appearance of the first nontrivial effect. At first, $D_{par}$ is practically identical to $D_{ang}$, but later $D_{par}$ starts to grow faster and diverges upon reaching the focal point. Afterwards, it becomes monotonically decreasing, with $D_{ang}$ eventually overtaking at the equidistance point. All this time both $D_{ang}$ and $\mu$ are monotonically increasing. The positions of the focal and equidistance points, which correspond to $\mu = 1$ and $\mu = 2$, are marked respectively by the solid and dashed lines.

Decreasing the impact parameter even more (Figure \ref{fig:fig1c}) reveals several more interesting effects. Again $D_{par}$ is initially growing faster than $D_{ang}$ and diverges at the focal point. However, $D_{ang}$ is not monotonic anymore and, together with $D_{par}$, vanishes at the conjugate point. From now on, both distances grow monotonically, with $D_{par}$ growing faster at first but later approaching a finite value.

A further decrease of the impact parameter (Figure \ref{fig:fig1d}) exposes only one new feature: the nonmonotonicity of $\mu$. We observe that there is a special point between two focal points, upon reaching which $\mu$ begins to decrease. Such behaviour is suggested by \eqref{eq:mudef1}, by which the value of $\mu$ at any focal point is equal to one. Therefore, at some point in between it has to become decreasing. Apart from that, we see a higher number of appearances of previously described features. 
$D_{ang}$ has an overall growing tendency, but it decreases to zero every time a conjugate point is passed. Analogously, $D_{par}$ diverges at focal points and vanishes at conjugate points.

In the faraway region the behaviour of distance measures becomes relatively simple. $D_{ang}$ grows without bounds, while $D_{par}$ approaches a constant value. $\mu$ also grows indefinitely, but its sign depends on the number of focal points passed.

Although these results represent only a few selected realisations of the problem, the qualitative properties survive in the general setting. We now present these properties in three different regimes.

\begin{figure}[H]
\centering
\includegraphics[width=1\linewidth]{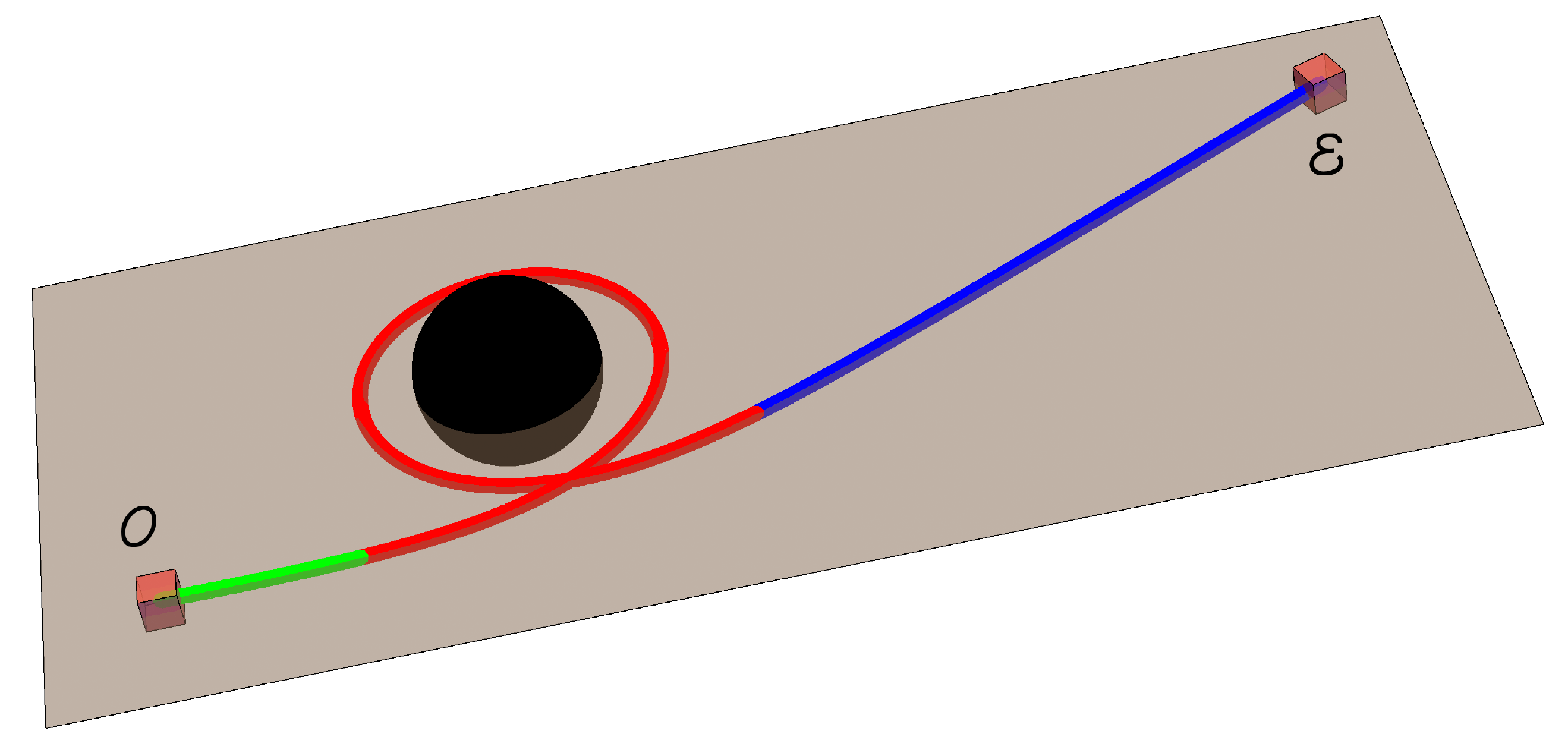}
\caption{
An illustration of the problem setting. The line depicts the null geodesic. The green, red and blue parts represent the initial, intermediate and faraway regions. The boxes at both endpoints stand for locally flat neighbourhoods. The black sphere marks the event horizon of the black hole.}
\end{figure}

\subsection{Initial region}

We begin with two types of bundles of rays: a vertex bundle and an initially parallel bundle. They can be understood by studying $\WXX$ and $\WXL$. In order to explain their behaviour in the initial region, we have to estimate the leading order behaviour. In the parallel propagated frame, expressing these operators as Taylor series around $\lambda_\calO = 0$ and using equations \eqref{eq:gdeM} to \eqref{eq:gdeic2} yields:

\begin{equation}
\begin{split}
{\WXL} \UD \mba \mbb &= \lambda \,\delta \UD \mba \mbb + \frac{\lambda^3}{3!} \calR \UD \mba \mbb  + \frac{\lambda^4}{4!} \paren{2 \dot \calR \UD 
\mba \mbb} + \frac{\lambda^5}{5!} \paren{3 \ddot \calR \UD \mba \mbb + \calR \UD \mba \mbc \calR \UD \mbc \mbb} + \map \calO {\lambda^6}\\
{\WXX} \UD \mba \mbb &= \delta \UD \mba \mbb + \frac{\lambda^2}{2!} \calR \UD \mba \mbb  + \frac{\lambda^3}{3!} \dot \calR \UD \mba \mbb  + \frac{\lambda^4}{4!}\paren{\ddot \calR \UD \mba \mbb + \calR \UD \mba \mbc \calR \UD \mbc \mbb} + \map \calO {\lambda^5}
\end{split}
\end{equation}
where $\dot \calR$ denotes the derivative of $\calR$ with respect to the affine parameter, and every curvature term is evaluated at $\lambda = 0$.

Spatial projections of these operators live in the 2-dimensional Euclidean space. Hence, one can apply the Cayley-Hamilton theorem to express the determinants in terms of traces:
\begin{equation}
{\det}_2 \paren {I + A} = 1 + \Tr A + \frac{\paren{\Tr A}^2 - \Tr \paren{A^2}}{2}
\end{equation}
This is particularly useful when one has to organize power expansions to higher orders. Also, it is instructive to decompose the optical tidal matrix as a sum of pure Ricci and Weyl terms, i.e. $\calR \UD \mba \mbb = -\frac 1 2 R_{\ell \ell} \delta \UD \mba \mbb  + C \UDDD \mba \ell \ell \mbb$. Then at $\calO$ the Taylor series expansion of determinants and $\mu$ has the following form:
\begin{equation}
\begin{split}
\det {\WXL} \UD \mba \mbb &= \lambda^2 \paren{1 - \frac{\lambda^2}{3!} R_{\ell\ell} - \frac{\lambda^3}{4!}2 \dot R_{\ell\ell} + \frac{\lambda^4}{5!} \sqbrk{\frac{5}{6}R_{\ell\ell}^2 - 3 \ddot R_{\ell\ell}} - \frac{\lambda^4}{5!} \frac{5}{3}C \UDDD \mba \ell \ell \mbb C \UDDD \mbb \ell \ell \mba} + \map \calO {\lambda^7}\\
\det {\WXX} \UD \mba \mbb &= 1 - \frac{\lambda^2}{2!} R_{\ell\ell} - \frac{\lambda^3}{3!} \dot R_{\ell\ell} + \frac{\lambda^4}{4!} \paren {2 R^2_{\ell\ell} - \ddot R_{\ell\ell} } - 2 \frac{\lambda^4}{4!} C \UDDD \mba \ell \ell \mbb C \UDDD \mbb \ell \ell \mba + \map \calO {\lambda^5}\\
\mu &= \frac{\lambda^2}{2!} R_{\ell\ell} + \frac{\lambda^3}{3!} \dot R_{\ell\ell} - \frac{\lambda^4}{4!} \paren {2 R^2_{\ell\ell} + \ddot R_{\ell\ell} } + 2 \frac{\lambda^4}{4!} C \UDDD \mba \ell \ell \mbb C \UDDD \mbb \ell \ell \mba + \map \calO {\lambda^5}
\end{split}
\end{equation}
Finally, we substitute these results into $\eqref{eq:dang}$ and $\eqref{eq:dpar}$ and obtain the leading order behaviour for the distance measures:
\begin{equation}
\begin{split}
\frac{D_{ang}}{\paren{\ell_\calO \cdot u_\calO}} &= \lambda - \frac{\lambda^3}{3!} \frac {R_{\ell\ell}} 2 - \frac{\lambda^4}{4!}\dot R_{\ell\ell} - \frac{\lambda^5}{5!} \frac 3 2 \ddot R_{\ell\ell} - \frac{\lambda^5}{5!} \frac 5 6 C\UDDD \mba \ell \ell \mbb C \UDDD \mbb \ell \ell \mba + \map \calO {\lambda^6}\\
\frac{D_{par}}{\paren{\ell_\calO \cdot u_\calO}} &= \lambda + \frac{\lambda^3}{3!}R_{\ell\ell} + \frac{\lambda^4}{4!} \dot R_{\ell\ell}+ \frac{\lambda^5}{5!} \paren{\frac {15} 4 R_{\ell\ell}^2 + \ddot R_{\ell\ell}} + \frac{\lambda^5}{5!} \frac {25} 6 C \UDDD \mba \ell \ell \mbb C \UDDD \mbb \ell \ell \mba + \map \calO {\lambda^6}
\end{split}
\end{equation}
From the above relations we conclude that $D_{ang}$, $D_{par}$ and $\mu$ are all regular for sufficiently short distances. Furthermore, whenever $R_{\ell\ell}$ is identically zero, e.g. in vacuum or in the presence of the cosmological constant, the difference between the operators, their determinants and derived distances is observable only at a relatively high order, which explains why in the beginning both vertex and initially parallel bundles are almost unaffected. In particular, Weyl contribution appears at the fifth order for distance measures and at the fourth order for $\mu$. On the other hand, when $R_{\ell\ell} > 0$, the difference is already visible at the third order for distances and at the second order for $\mu$. Moreover, in the leading order, $D_{par}$ is larger than $D_{ang}$, and $\mu$ is positive. In fact, it is possible to present a more general, non-perturbative statement, valid in any spacetime. In the companion paper \cite{serbenta2021} we prove the following result:
\begin{theorem}
Let $\calO$ and $\calE$ be two points along a null geodesic $\gamma$ such that $\calO$ lies in the causal future of $\calE$ and let the NEC hold along $\gamma_0$ between $\calO$ and $\calE$. Assume also that between $\calO$ and $\calE$ there are no singular points of the infinitesimal bundle of rays parallel at $\calO$. Then we have
\begin{equation}
 \mu \ge 0.
\end{equation}
Moreover, $\mu=0$ iff the transverse optical tidal tensor $R\UD{\bm A}{\bm \mu \bm \nu \bm B}\,\ell^{\bm \mu}\,\ell^{\bm \nu}$ vanishes along $\gamma_0$ between $\calO$ and $\calE$.
\end{theorem}

\subsection{Intermediate region}

In the intermediate region geodesics may circle the black hole, but eventually they escape to infinity unless they fall onto the photon sphere or the black hole. In any case, both the initially parallel and vertex bundles start to converge due to Weyl focusing, as can be seen from \eqref{eq:raye} and \eqref{eq:rays}. The parallel bundle is the first one to be focused because already at $\lambda_\calO$  because its expansion is zero, and later it can only decrease. The position of the corresponding focal point depends on the parameters of the null geodesic, but it is reached sooner than the corresponding conjugate point. At the focal point $D_{par}$ diverges and $\mu = 1$ while $D_{ang}$ is regular. 

If, after passing the focal point, the geodesic is still sufficiently close to the black hole, it will encounter the conjugate point where the vertex bundle will converge back to a point. Even though the initial conditions determine the position of this point, the angular coordinate $\phi$ is actually independent of them and always equals a multiple of $\pi$ (assuming that initially $\phi = 0$). At the conjugate points both $D_{par}$ and $D_{ang}$ vanish, but $\mu$ is regular.

By looking at the expressions for BGOs in the parallel propagated SNT one can notice that the dependence on $\phi$ is periodic with the period of $2\pi$, while the angular coordinates of focal and conjugate points differ by a multiple of $\pi$. This can be easily understood from the symmetry of the problem. In a spherically symmetric spacetime every perturbed geodesic is contained in a plane. All these planes are tilted with respect to each other, but they share one common line. All the points from which geodesics emanate or to which they converge lie on this line. Furthermore, the geodesic equation depends on the square of angular momentum. Tilting a plane implies a perturbation of the angular momentum, i.e. $L_x^2 \to L_x^2 + 2 L_x \delta L_x + \delta L_x^2$. However, in the domain of the linear geodesic deviation, only linear terms should be considered. In addition to this, both $L_x$ and $L_y$ vanish in the aligned coordinates, which implies that the linear perturbation is also absent. Therefore, every ray of the null congruence satisfies the same geodesic equation. Provided we choose the correct affine parameter gauge, they meet at the same point.

\begin{figure}[H]
\centering
\includegraphics[width=0.5\linewidth]{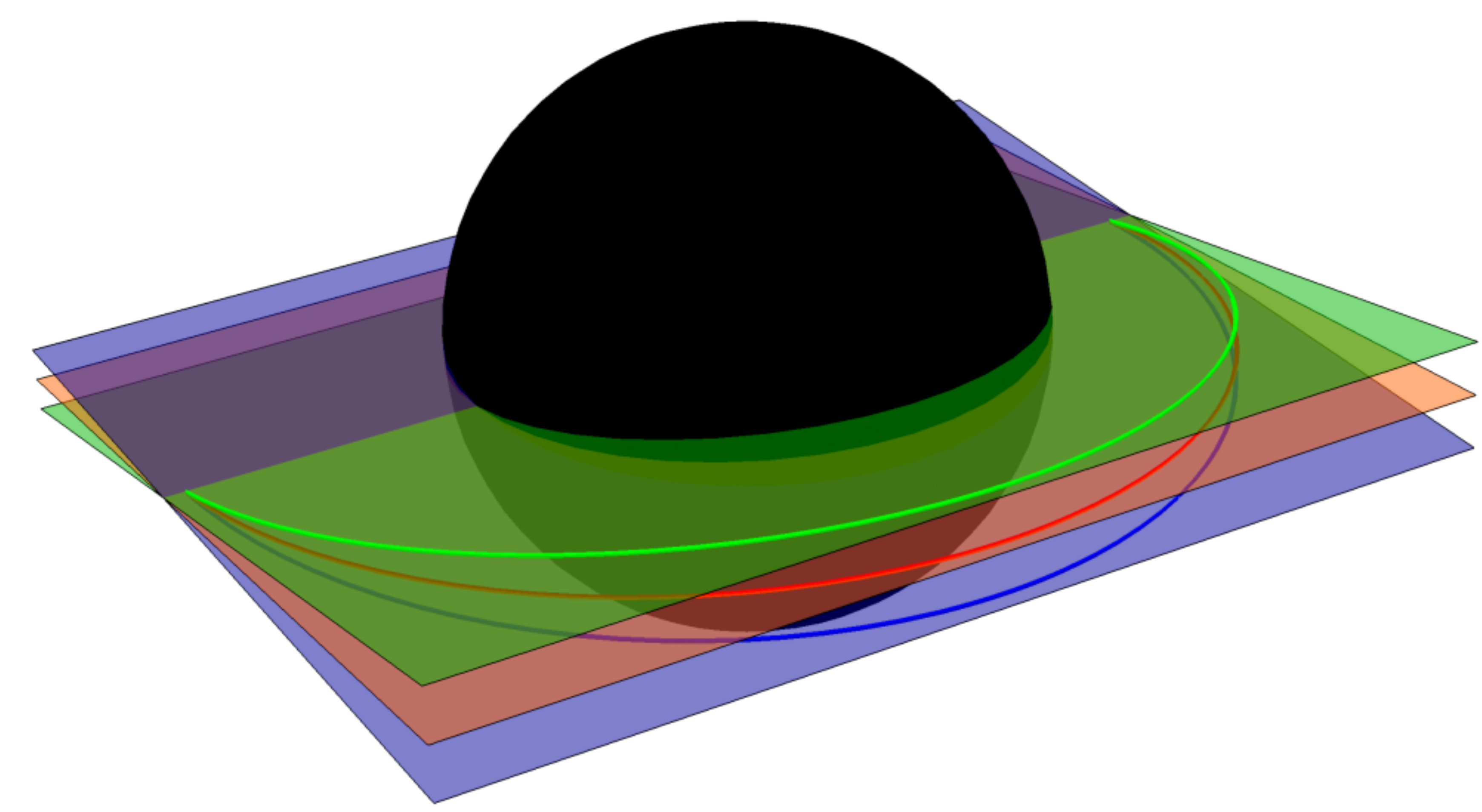}
\caption{
Due to the symmetries of Schwarzschild spacetime every geodesic is completely contained in a plane passing through the center of the black hole. Geodesics that emerge from the same point but differ in their vertical alignments belong to different planes which share a line passing through the initial point and the center. All points conjugate to the initial one lie on this line.}
\end{figure}

After the focal point one can almost always expect an equidistant point where $D_{par} = D_{ang}$. This point may come either before or after the conjugate point. At this point $\mu$ is either 0 or 2, and all observables are regular.

One has to note that the number of occurrences of these special points depends on the total azimuthal angle swept by the geodesic, measured by $\phi$. It is determined by the parameters of the geodesic, i.e. $b$ and $r_\calO$. Even if we take $\lambda$ from $\lambda_{\cal O}$ up to infinity, the range of $\phi$ is bounded for geodesics that are not trapped by the black hole. It may therefore happen that several sequences of focal, conjugate or equidistant points will be traversed, or that in the end the counts of each type of point will be different. However, the qualitative behaviour at these points and in between is the same.
  
\subsection{Faraway region}

In the faraway region the curvature is becoming arbitrarily small, and the geodesic approaches a radial null line. This results in the Weyl focusing becomming negligible. Thus, effectively, nearby light rays propagate as if they were in the Minkowski spacetime. From (\ref{eq:gdeM}) we have  in that case:
\begin{equation}
\begin{split}
\WXX &\sim A_{XX} + \lambda\,B_{XX} \\
\WXL &\sim A_{XL} + \lambda\,B_{XL} 
\end{split},
\end{equation}
where $A_{XX}, A_{XL},B_{XX}, B_{XL}$ are constant matrices. Their precise form depends on the whole history of the null geodesics from the observation point up to the faraway region.  Taking their determinants yields the asymptotic behaviour for $\lambda \to \infty$:
\begin{equation}
\begin{split}
\det {\WXX} \UD \mba \mbb &\sim \paren {\det {B_{XX}} \UD \mba \mbb} \lambda^2\\
\det {\WXL} \UD \mba \mbb &\sim  \paren{\det {B_{XL}} \UD \mba \mbb}\lambda^2
\end{split}
\end{equation}
In a generic situation we may assume that these determinants do not vanish. Then, according to $\eqref{eq:dang}$ and $\eqref{eq:dpar}$, $D_{ang}$ and $D_{par}$ have the following asymptotic behaviour:
\begin{equation}
\begin{split}
D_{ang} &\sim \lambda\\
D_{par} &\sim \const.
\end{split}
\end{equation}
In other words, sufficiently far away $D_{ang}$ is almost a linear function, while $D_{par}$ approaches a constant value. From the intermediate region analysis we know that $D_{par}$ is initially larger than $D_{ang}$. Therefore, in order to shrink to a fixed value, it must at some point match $D_{ang}$. For this reason the existence of the equidistant point is guaranteed for a generic geodesic.

The constancy of $D_{par}$ is surprising, but at the same time it is actually a rather generic feature of asymptotically flat spacetimes. In Minkowski spacetime, for a baseline of fixed length, the parallax angle depends only on the position of the source. The further lies the source, the lesser the parallax angle, and this angle is close to zero for infinitely distant objects. In the general case, however, the trajectory of light will pass through a curved region and will be deflected. Then the total parallax is the sum of parallax in relatively flat regions and the contribution of light deflection in-between. Even if the parallax in the outer region can be made arbitrarily small, the passing of light through a curved region leaves its imprint that does not go away. The gravitational lensing best illustrates this: the position of an apparent image is very sensitive to its proximity to the 2-dimensional projection of the lensing body and its parallax effect. The parallax of a nearby lens, such as the Schwarzschild black hole in our case,  combined with light deflection  provides this way an additional parallax effect for very distant objects. This in turn limits the effective parallax distance to these object, as measured by the observer.

It may happen that for some initial conditions that matrices  $B_{XX}$ and $B_{XL}$ are degenerate. In that case the asymptotic analysis of the behaviour of $D_{ang}$ and $D_{par}$ does not apply.  For example, in the Schwarzschild spacetime, radial null geodesics correspond to the principal null directions and are shear free. The behaviour of distances then is analogous to the one in flat space where both $D_{ang}$ and $D_{par}$ grow linearly and are  both equal all along the geodesic. However, these situations require extreme fine-tuning of the initial data  and do not represent the generic behaviour.

\section{Conclusion}

In this paper, we have presented two exact methods of solving the geodesic deviation equation and deriving the bilocal geodesic operators. The first method is based on the linear variation of the solution to the geodesic equation with respect to initial data.  It requires an expression for the general solution of the geodesic equation in an explicit form or at least a sufficient number of implicit relations defining the geodesic.  The second method makes use of the conserved quantities generated by Killing vectors. Every such quantity generates a conservation law for the geodesic deviation equation. In both cases, BGO's can be read off by taking partial derivatives of exact solutions or variations with respect to covariant perturbations of the initial data. Then we apply both methods to obtain the BGOs in a 4-dimensional static spherically symmetric spacetime. In these spacetimes, a generic null geodesic is always contained in a plane passing through the origin. This allows us to reduce the dimensionality of the problem. Finally, to isolate physical effects, we project the BGOs onto a parallel-transported SNT.

In the second part of the paper, we investigated the behaviour of distance measures such as the angular diameter distance, the parallax distance and the distance slip in the Schwarzschild spacetime. We considered cases where both the source and the observer are located outside of the photon sphere. In the numerical study, we considered trajectories with a static observer and four different impact parameters. We have noticed that as the impact parameter approaches the value corresponding to the photon sphere, the occurrence and the strength of various nontrivial optical effects increases. One can observe the formation of various special points where the parallax distance diverges, parallax and angular distances become zero or equal to each other. Another interesting feature of this spacetime is that the parallax distance of a source positioned infinitely far away is finite.

In the last part, we provide a more general explanation for the observed phenomena. In the absence of matter, the curvature effects appear at a relatively high order, which explains why initially the distances are almost the same. In the intermediate region, the light is refocused to a line, and the number of such events depends on the total deflection angle. All points where this focusing happens lie on a line in the geodesic plane, which passes through the center of the black hole. In the faraway region, light rays propagate in effectively flat spacetime. However, at the same time, they carry the imprint of the previous regimes. We then show that in the faraway region the generic behaviour of distance measures in the leading order is linear for the angular diameter distance and constant for the parallax distance.

\section*{Acknowledgments}

The work was supported by the National Science Centre, Poland (NCN) via the SONATA BIS programme, Grant No.~2016/22/E/ST9/00578 for the project
\emph{``Local relativistic perturbative framework in hydrodynamics and general relativity and its application to cosmology''}. 

\appendix
\section*{Appendix}

\subsection{Variations of the implicit solutions of the geodesic equation} \label{app:implicitvar}

\subsubsection{Variation of conserved quantities}

We want to find how variations of initial position and direction affect \eqref{eq:kil0E}-\eqref{eq:norm}. Due to the assumed symmetry, we can always choose aligned coordinates, where $L_x$ and $L_y$ are zero. However, this does not have to hold for their variations. The results are:
\begin{equation}
    \begin{split}
        \delta L_x \big |_\calO &= C_\calO \, \ell^\phi_\calO \, \delta \theta_\calO\\
        \delta L_y \big |_\calO &= C_\calO \, \delta \ell^\theta_\calO\\
        \delta L_z \big |_\calO &= \frac{C_\calO'}{C_\calO}L_z \delta r_\calO + C_\calO \delta \ell^\phi_\calO\\
        \delta E \big |_\calO &= -A_\calO \delta \ell^t_\calO + E \frac{A'_\calO}{A_\calO}\delta r_\calO\\
        \delta \epsilon \big \vert_\calO &= \sqbrk{-A_\calO' \frac{E^2}{A^2_\calO} + B_\calO' \paren{\ell^r_\calO}^2 + C_\calO' \frac{L^2}{C_\calO^2}}\delta r_\calO + 2E \delta \ell^t_\calO + 2 B_\calO \ell^r_\calO \delta \ell^r_\calO + 2 L_z \delta \ell^\phi_\calO
    \end{split}
\end{equation}
Note that the variation of $L_x$ is simply proportional to the variation of $\theta_\calO$, and we may safely substitute it everywhere by $\delta \theta_\calO$. The remaining four variations of 
conserved quantities can be used to parametrize the variations of the four components of the initial four-momentum $\ell_\calO^\mu$.
We also point out that even though $\epsilon$ here is arbitrary, eventually, we will set $\epsilon = 0$ to limit ourselves to null geodesics.

\subsubsection{Variation of the implicit equations} \label{app:varimplicit}

Now we switch to the variation of solutions to the geodesic equations. We begin with \eqref{eq:gth} which we vary and then evaluate in the aligned coordinates:
\begin{equation}
    \delta \theta = \cos \phi \delta \theta_\calO + \sin \phi \frac{\delta \ell^\theta_\calO}{\ell^\phi_\calO}
\end{equation}
Next we use \eqref{eq:kil0Lx} and \eqref{eq:kil0Ly} to obtain $\ell^\theta$:
\begin{equation}
    \ell^\theta = \frac{L_y \cos \phi - L_x \sin \phi}{C}
\end{equation}
In the aligned coordinates its variation yields:
\begin{equation}
    \delta \ell^\theta = \frac{C_\calO}{C} \paren{\cos \phi \,\delta \ell^\theta_\calO - \sin \phi \, \ell^\phi_\calO \, \delta \theta_\calO}
\end{equation}
Note that variations of $\theta$ and $\ell^\theta$ decouple from variations of other components of the geodesic. This is a consequence of the existence of a plane containing the geodesic.

Next we will consider the variation of $r$. We choose $\lambda$ to be our dependent variable to avoid working with the formal solution $\map r \lambda$. From the normalization condition \eqref{eq:norm} we get:
\begin{equation}
    \label{eq:rint}
    \lambda - \lambda_\calO = \fint_{r_\calO}^r \pm_r \sqrt{\frac{ABC}{AC\epsilon + E^2 C - L^2 A}} d \tilde r
\end{equation}
From now on we set $\lambda_\calO = 0$ for convenience. We have that:
\begin{equation}
    \delta \lambda = \frac{\delta r}{\ell^r} - \frac{\delta r_\calO}{\ell^r_\calO} + L I_{BC} \delta L - E I_{AB} \delta E - \frac{I_B}{2} \delta \epsilon
\end{equation}
This can be easily solved for $\delta r$:
\begin{equation}
    \delta r = \ell^r \paren{ \delta \lambda + \frac{\delta r_\calO}{\ell^r_\calO} - I_{BC} L \delta L + I_{AB} E \delta E + \frac{I_B}{2}  \delta \epsilon}
\end{equation}
Variation of $\ell^r$ is straightforward:
\begin{equation}
    \delta \ell^r = \frac{1}{\ell^r} \paren{\frac{\delta \epsilon}{2B} + \frac{E \delta E}{AB} - \frac{L \delta L}{BC}} - \frac{1}{\ell^r} \paren{\frac{\epsilon B'}{B^2} + \frac{E^2}{AB} \paren{\frac{A'}{A} + \frac{B'}{B}} - \frac{L^2}{BC} \paren{\frac{B'}{B} + \frac{C'}{C}}} \delta r,
\end{equation}
the prime denoting here $A'(r) = A_{,r}$ etc.

Similarly, for $t$ and $\ell^t$ we have:

\begin{equation}
\begin{split}
    \delta t &= \delta t_\calO + E \paren{\frac{\delta r_\calO}{A_\calO \ell^r_\calO} - \frac{\delta r}{A \ell^r} } + I_{ABC} L\paren{L \delta E - E\delta L} + I_{AB} \paren{\frac E 2 \delta \epsilon - \epsilon \delta E}\\
    \delta \ell^t &= - \frac{\delta E}{A} + \frac{E}{A^2}A' \delta r
    \end{split}
\end{equation}
In order to vary $\phi$ and $\ell^\phi$ we start from \eqref{eq:implicitphi} and \eqref{eq:kil0Lz}. Here we have to recall that by $\eqref{eq:gth}$, $\theta$ depends on $\phi$ as well. However, in the aligned coordinates, $\theta$ variations simply drop out, and we are left with the standard result:

\begin{equation}
    \begin{split}
    \delta \phi &= \delta \phi_\calO + L_z \paren{\frac{\delta r}{C \ell^r} - \frac{\delta r_\calO}{C_\calO \ell^r_\calO}} + I_{ABC} E \paren{E \delta L_z - L_z \delta E} + I_{BC} \paren{\epsilon \delta L_z - \frac{L_z \delta \epsilon}{2}}\\
    \delta \ell^\phi &= \frac{\delta L_z}{C} - \frac{L_z}{C^2} C' \delta r
    \end{split}
\end{equation}
In order to find $\calW$ operators, we use variations we have obtained so far together with $\eqref{eq:covW}$. Firstly, by reading off components for each variation, we obtain functions corresponding to partial derivatives in $\eqref{eq:covW}$. Then, we calculate Christoffel symbols and vectors tangent to the geodesic and evaluate them at one of the endpoints as prescribed by $\eqref{eq:covW}$.

\subsection{Solution of GDE}

Here we write down the solution of GDE using the method of conserved quantities.

\begin{equation*}
    \begin{split}
        \xi^\theta &= \kappa_1 \sin \phi + \frac{\Sigma_x}{L_z} \cos \phi\\
    \nabla_\ell \xi^\theta &= \xi^\theta \paren{\cot \phi \, \ell^\phi + \frac{C'}{2C} \ell^r} - \frac{\Sigma_x}{C \sin \phi}\\ 
        \xi^r &=\ell^r \paren {\kappa_2 - E\, \Sigma_T I_{AB} + L_z \Sigma_z I_{BC} + \mathcal B I_B}\\
        \nabla_\ell \xi^r &= \frac{\mathcal B - E \nabla_\ell \xi^t - L_z \nabla_\ell \xi^\phi}{B \ell^r}\\
        \xi^t &= \kappa_3 + \kappa_2 E \paren{\frac 1 {A_\calO} - \frac 1 A} + \Sigma_T \paren{E^2 \frac{I_{AB}} A - L_z^2 I_{ABC} + \epsilon I_{AB}} + L_z \Sigma_z E \paren{ I_{ABC} - \frac{I_{BC}}{A} } + E \mathcal B \paren{I_{AB} - \frac{I_B}{A}}\\
        \nabla_\ell \xi^t &= \frac{A' \ell^r}{2A} \paren{\kappa_3 + \kappa_2 \frac{E}{A_\calO} + L_z I_{ABC} \paren{\Sigma_z E - \Sigma_T L_z} + I_{AB} \paren{\Sigma_T \epsilon + E \mathcal B} } + \frac {\Sigma_T} A\\
    \xi^\phi &= \kappa_4 + \kappa_2 L_z \paren{\frac 1 C - \frac 1 {C_\calO}} + \Sigma_z \paren{L_z^2 \frac{I_{BC}}{C} - E^2 I_{ABC}  - \epsilon I_{BC}} + E \Sigma_T L_z \paren{I_{ABC} - \frac{I_{AB}}{C} } + L_z \mathcal B \paren{\frac{I_B}{C} - I_{BC}}
\end{split}
\end{equation*}

\begin{equation*}
\begin{split}
    \nabla_\ell \xi^\phi &= \frac{C' \ell^r}{2C} \paren{E I_{ABC} \paren{\Sigma_T L_z - \Sigma_z E} - I_{BC} \paren{\Sigma_z \epsilon + L_z \mathcal B} + \kappa_4 - \kappa_2\frac{L_z}{C_\calO}} - \frac{\Sigma_z} C
    \end{split}
\end{equation*}

\subsection{BGO's expressed in the coordinate tetrad}
\label{appendix:BGOc}

\begin{equation*}
    \begin{split}
        {W_{XX}} \UD t t &= 1 + \frac 1 2 A'_\calO \ell^r_\calO \paren{L^2_z I_{ABC} - \frac{E^2}{A} I_{AB} - \epsilon I_{AB}} \quad {W_{XX}} \UD t \phi = \frac{C'_\calO \ell^r_\calO}{2} E L_z \paren{I_{ABC} - \frac {I_{BC}} A}\\
        {W_{XX}} \UD t r &= \frac{E}{\ell^r_\calO} \paren{\frac 1 {A_\calO} - \frac 1 A} + \frac{C'_\calO L^2_z E}{2 C_\calO} \paren{\frac {I_{BC}} A - I_{ABC}} + \frac{A'_\calO E}{2 A_\calO} \paren{L^2_z I_{ABC} - \frac{E^2}{A} I_{AB}  - \epsilon I_{AB}}\\
        {W_{XX}} \UD r t &= \frac{A'_\calO \ell^r_\calO} 2  E \ell^r I_{AB} \quad {W_{XX}} \UD r \phi = \frac{C'_\calO \ell^r_\calO}{2} L_z  \ell^r I_{BC} \\ {W_{XX}} \UD r r &= \frac{\ell^r}{\ell^r_\calO} + \frac{\ell^r}{2} \paren{E^2 \frac{A'_\calO}{A_\calO} I_{AB} - L^2_z \frac{C'_\calO}{C_\calO} I_{BC}} \quad
        {W_{XX}} \UD \theta \theta = \cos \phi - C'_\calO \ell^r_\calO \frac{\sin \phi}{2 L_z} \\
        {W_{XX}} \UD \phi t &= \frac{A'_\calO \ell^r_\calO}{2} L_z E \paren{\frac {I_{AB}} C - I_{ABC}} \quad {W_{XX}} \UD \phi \phi = 1 + \frac{C'_\calO \ell^r_\calO}{2} \paren{ \frac {L^2_z} C I_{BC} - E^2 I_{ABC} - \epsilon I_{BC} }\\ {W_{XX}} \UD \phi r &= \frac{L_z}{\ell^r_\calO} \paren {\frac{1}{C} - \frac{1}{C_\calO}} + \frac{A'_\calO}{2 A_\calO} E^2 L_z \paren {\frac {I_{AB}} C - I_{ABC}} + \frac{C'_\calO}{2C_\calO} L_z \paren {E^2 I_{ABC} - \frac{L^2_z}{C} I_{BC} + \epsilon I_{BC}}\\
         {W_{XL}} \UD t t &= E^2 \paren{I_{AB} - \frac{I_B}{A}} + A_\calO \paren {E^2 \frac{I_{AB}}{A} - L^2_z I_{ABC} + \epsilon I_{AB}} \quad {W_{XL}} \UD t r = E B_\calO \ell^r_\calO \paren{I_{AB} - \frac{I_B}{A}}\\
        {W_{XL}} \UD t \phi &= E L_z \paren{I_{AB} - \frac{I_B}{A} + C_\calO \paren{\frac{I_{BC}}{A} - I_{ABC} }} \quad {W_{XL}} \UD \theta \theta = \frac{C_\calO}{L_z} \sin \phi\\
        {W_{XL}} \UD r  t &= E \ell^r \paren{I_B - A_\calO I_{AB} }\quad {W_{XL}} \UD r r = B_\calO \ell^r_\calO \ell^r I_B \quad {W_{XL}} \UD r \phi = L_z \ell^r \paren{I_B - C_\calO I_{BC}} \\
        {W_{XL}} \UD \phi t &= E L_z \paren{ A_\calO I_{ABC} - I_{BC} + \frac {I_B - A_\calO I_{AB}} C } \quad {W_{XL}} \UD \phi r = B_\calO \ell^r_\calO L_z \paren{\frac {I_B} C - I_{BC}}\\ {W_{XL}} \UD \phi \phi &= L^2_z \paren{\frac {I_B} C - I_{BC}} + C_\calO \paren {E^2 I_{ABC} - \frac {L^2_z} C I_{BC} + \epsilon I_{BC}}\\
        {W_{LX}} \UD t t &= \frac 1 {2A} \paren{\ell^r A' - \ell^r_\calO A'_\calO + \frac{A'_\calO \ell^r_\calO A' \ell^r}{2}\paren{L^2_z I_{ABC} - \epsilon I_{AB}} } \quad {W_{LX}} \UD t \phi = \frac{L_z E} {4A} A' \ell^r C'_\calO \ell^r_\calO I_{ABC}\\
        {W_{LX}} \UD t r &= \frac{E}{2 A A_\calO \ell^r_\calO} \paren{A' \ell^r - A'_\calO \ell^r_\calO} + \frac{E L^2_z}{4 A} A' \ell^r \paren{\frac{A'_\calO}{A_\calO} - \frac{C'_\calO}{C_\calO} } I_{ABC} - \frac{A' \ell^r I_{AB}\epsilon A'_\calO E}{4A A_\calO}\\
        {W_{LX}} \UD r t &= \frac{E}{2AB \ell^r} \paren{A'_\calO \ell^r_\calO - A' \ell^r} + \frac{E L^2_z}{4B} A'_\calO \ell^r_\calO \paren{\frac {C'} C - \frac {A'} A} I_{ABC} + \frac{A'_\calO \ell^r_\calO A' E \epsilon}{4AB} I_{AB}\\
        {W_{LX}} \UD r r &= \frac{E^2 \paren{A'_\calO \ell^r_\calO - A' \ell^r}}{2ABA_\calO \ell^r \ell^r_\calO}  + \frac{L^2_z \paren{C' \ell^r - C'_\calO \ell^r_\calO}}{2BC C_\calO \ell^r \ell^r_\calO} + \frac{E^2 L^2_z}{4B} \paren{\frac{A'}{A} - \frac{C'}{C}} \paren{\frac{C'_\calO}{C_\calO} - \frac{A'_\calO}{A_\calO}} I_{ABC} \\&+ \frac{\epsilon}{4B} \paren{E^2 \frac{A'_\calO A'}{A_\calO A} I_{AB} - L_z^2 I_{BC} \frac{C'_\calO C'}{C_\calO C}}\\
        {W_{LX}} \UD r \phi &= \frac{L_z}{2BC \ell^r} \paren{C'_\calO \ell^r_\calO - C' \ell^r} + \frac{E^2 L_z}{4B} C'_\calO \ell^r_\calO \paren{\frac{C'}{C} - \frac{A'}{A}} I_{ABC} + \epsilon \frac{L_z C'}{4BC} C'_\calO \ell^r_\calO I_{BC} \\
        {W_{LX}} \UD \theta \theta &= \frac{\cos \phi}{2C} \paren{C' \ell^r - C_\calO' \ell^r_\calO} - \paren {\frac {L_z} C + \frac{C' \ell^r C'_\calO \ell^r_\calO}{4CL_z}} \sin \phi\\
        {W_{LX}} \UD \phi t &= -\frac{L_z E}{4C} C' \ell^r A_\calO' \ell^r_\calO I_{ABC} \quad {W_{LX}} \UD \phi r = \frac{L_z \paren{C'_\calO \ell^r_\calO - C' \ell^r}}{2C C_\calO \ell^r_\calO} + \frac{L_z E^2}{4C} C' \ell^r \paren{\frac{C'_\calO}{C_\calO} - \frac{A'_\calO}{A_\calO}} I_{ABC} + \frac{C'_\calO L_z C' \ell^r}{4 C C_\calO} \epsilon I_{BC}\\
        {W_{LX}} \UD \phi \phi &= \frac{1}{2C} \paren{C' \ell^r - C'_\calO \ell^r_\calO} - \frac{E^2}{4C} C' \ell^r C'_\calO \ell^r_\calO I_{ABC} - \frac{\epsilon I_{BC}}{4C}C' \ell^r C'_\calO \ell^r_\calO \quad
        {W_{LL}} \UD t r = \frac{E}{2A} A' \ell^r B_\calO \ell^r_\calO I_{AB}\\ {W_{LL}} \UD t t &= \frac{A_\calO}{A} + \frac{A' \ell^r}{2A} \paren{E^2 I_{AB} - L^2_z A_\calO I_{ABC} + \epsilon I_{AB} A_\calO} \quad {W_{LL}} \UD t \phi = \frac{E L_z}{2A} A' \ell^r \paren{I_{AB} - C_\calO I_{ABC}}
\end{split}
\end{equation*}
\begin{equation*}
\begin{split}
        {W_{LL}} \UD r t &= \frac{E \paren{A - A_\calO}}{AB \ell^r} + \frac{E}{2B} \paren{L^2_z \frac{C'}{C} I_{BC} - E^2 \frac{A'}{A} I_{AB}} + \frac{L^2_z E A_\calO}{2B} \paren{\frac{A'}{A} - \frac{C'}{C}} I_{ABC} - \epsilon \frac{A'A_\calO E}{2AB} I_{AB}\\
        {W_{LL}} \UD r r &= \frac{B_\calO \ell^r_\calO}{B \ell^r} + \frac{B_\calO \ell^r_\calO}{2B} \paren{L^2_z \frac{C'}{C} I_{BC} - E^2 \frac{A'}{A} I_{AB}} \\
        {W_{LL}} \UD r \phi &= \frac{L_z \paren{C - C_\calO}}{BC \ell^r} + \frac{L_z}{2B}\paren{L^2_z \frac{C'}{C} I_{BC}- E^2 \frac{A'}{A} I_{AB}} + \frac{E^2 L_z C_\calO}{2B} \paren{\frac{A'}{A} - \frac{C'}{C}} I_{ABC} - \epsilon \frac{L_z C' C_\calO}{2BC} I_{BC}\\
        {W_{LL}} \UD \theta \theta &= \frac{C_\calO}{C} \cos \phi + \frac{C_\calO}{2CL_z} C' \ell^r \sin \phi\quad {W_{LL}} \UD \phi t = \frac{E L_z}{2C} C' \ell^r \paren{A_\calO I_{ABC} - I_{BC}}\\
        {W_{LL}} \UD \phi r &= - \frac{L_z B_\calO \ell^r_\calO}{2C} C' \ell^r I_{BC} \quad {W_{LL}} \UD \phi \phi = \frac{C_\calO}{C} + \frac{C' \ell^r}{2C} \paren{E^2 C_\calO I_{ABC} - L^2_z I_{BC} + \epsilon I_{BC} C_\calO}
    \end{split}
\end{equation*}

\subsection{Optical tidal matrix in aligned coordinate tetrad}

\begin{equation}
\begin{split}
    \calR \UD t t &= \frac{E^2 + \epsilon A}{E B \ell^r} \calR \UD t r + \paren{\frac{L_z}{2ABC}}^2 \paren{A'C \paren{AB}' + AB \paren{A' C' - 2 A'' C}} \quad \calR \UD t r = \frac{\paren{2 A B A'' - A' \paren{AB}' } E \ell^r}{4 A^3 B} \\\calR \UD t \phi &= \frac{E L_z A' C'}{4 A^2 B C} \quad
    \calR \UD r t = - \frac{A}{B} \calR \UD t r \quad \calR \UD r r = -\frac{E}{B \ell^r} \calR \UD t r - \frac{L_z}{B \ell^r} \calR \UD \phi r \quad \calR \UD r \phi = \frac{C}{B} \calR \UD \phi r \quad
    \calR \UD \phi t = - \frac{A}{C} \calR \UD t \phi \\ \calR \UD \phi r &= \frac{L_z \ell^r}{4BC^2} \paren{B' C' + \frac B C \paren{C'^2 - 2 C C''}} \quad \calR \UD \theta \theta = \calR \UD \phi \phi + \frac{L_z^2}{4BC^3} \paren{C'^2 - 4BC}\\
    \calR \UD \phi \phi &= - \paren{\frac{E}{2ABC}}^2 \paren {C'C \paren{AB}' + AB \paren{C'^2 - 2CC''}} + \frac{L_z^2 - \epsilon C}{L_z B \ell^r} \calR \UD \phi r
\end{split}
\end{equation}

\subsection{Optical tidal matrix in the semi-null tetrad $(\epsilon = 0)$}

\begin{equation}
\begin{split}
    \calR \UDb 1 1 &= \paren{\frac{L_z}{2 B C}}^2 \paren{B \paren{C''-4B} + \paren{BC'}' + \frac{2B}{C}\paren{C'^2 - 2CC''}} -\paren{\frac{E}{2 A B C}}^2 \paren{AB \paren{C'^2 - 2C C''} + C C' \paren{AB}'}\\
    \calR \UDb 2 2 &= \paren{\frac{L_z}{2ABC}}^2\paren{A'C \paren{AB}' + AB \paren{A'C' - 2CA''}} - \paren{\frac{E}{2ABC}}^2 \paren{AB\paren{C'^2 - 2CC''} + CC' \paren{AB}'}\\
    \calR \UDb 2 0 &= \Psi \calR \UDb 2 2 - \frac{E \ell^r Q}{4 \paren{ABC}^{\frac 3 2} L_z} \paren{AB \paren{C'^2 - 2CC''} + CC' \paren{AB}'} \quad \calR \UDb 3 2 = \frac {\calR \UDb 2 0} Q \\ \calR \UDb 3 0 &= \frac{Q}{\paren{2BC}^2} \paren{B\paren{C'^2 - 2CC''} + CC'B'} - \paren{\frac{E}{\ell^r L_z} \sqrt{\frac{C}{AB}} + 2 \frac \Psi Q} \paren{\Psi \calR \UDb 2 2 - \calR \UDb 2 0} + \frac{\Psi^2}{Q} \calR \UDb 2 2
\end{split}
\end{equation}

\subsection{BGO's in the semi-null tetrad}
\label{appendix:BGOs}
\begin{equation}
    \begin{split}
    {\WXL} \UDb 0 0 &= {\WXL} \UDb 3 3 = \lambda \quad {\WXL} \UDb 1 1 = \frac{\sqrt{C C_\calO}}{L_z} \sin \phi \quad {\WXL} \UDb 2 2 = \sqrt{A A_\calO B B_\calO C C_\calO} \ell^r \ell^r_\calO I_{ABC} \\
    {\WXL} \UDb 2 0 &=  \frac{E Q}{L_z} \ell^r \sqrt{ABC} \paren{C_\calO I_{ABC} - I_{AB}} - \lambda \Psi \quad {\WXL} \UDb 3 2 = \frac{E}{L_z} \ell^r_\calO \sqrt{A_\calO B_\calO C_\calO} \paren{C I_{ABC} - I_{AB}} + {\WXL} \UDb 2 2 \frac{\Psi}{Q}\\
    {\WXL} \UDb 3 0 &= \frac{Q}{L_z^2} \paren{E^2 \paren{CC_\calO I_{ABC} - \frac{C + C_\calO}{2}I_{AB}} + L_z^2 \paren{I_B - \frac{C + C_\calO}{2}I_{BC}}} + \frac{\Psi}{Q} {\WXL} \UDb 2 0 + \frac{\lambda \Psi^2}{2Q}
    \end{split}
\end{equation}

\begin{equation}
    \begin{split}
        {\WXX} \UDb 0 0 &= {\WXX} \UDb 3 3 = 1 \quad {\WXX} \UDb 1 1 = \sqrt{\frac{C}{C_\calO}} \paren{\cos \phi - \frac{C'_\calO \ell^r_\calO \sin \phi}{2L_z}} \\
        {\WXX} \UDb 2 0 &= -\sqrt{ABC}\frac{C'_\calO \ell^r_\calO \ell^r E Q}{2L_z} I_{ABC} -\Psi \quad {\WXX} \UDb 2 2 = \sqrt{\frac{ABC}{A_\calO B_\calO C_\calO}}\paren{\frac{\ell^r}{\ell^r_\calO} + \frac{\ell^r}{2} \paren{L_z^2 A'_\calO - E^2 C'_\calO} I_{ABC}} \\
        {\WXX} \UDb 3 2 &= \frac{E}{2 L_z \sqrt{A_\calO B_\calO C_\calO}} \paren{C I_{ABC} \paren{L_z^2 A'_\calO - E^2 C'_\calO} + L_z^2 \paren{C'_\calO I_{BC} - A'_\calO I_{AB}} + \frac{2\paren{C - C_\calO}}{\ell^r_\calO}} + \frac{\Psi}{Q} {\WXX} \UDb 2 2\\
        {\WXX} \UDb 3 0 &= \frac{Q}{2L^2_z} \paren{C - C_\calO + C'_\calO \ell^r_\calO \paren{I_{BC} L^2_z - E^2 I_{ABC}C}} + {\WXX} \UDb 2 0 \frac \Psi Q + \frac{\Psi^2}{2Q}
    \end{split}
\end{equation}

\begin{equation}
    \begin{split}
     {\WLL} \UDb 0 0 &= {\WLL} \UDb 3 3 = 1 \quad {\WLL} \UDb 1 1 = \sqrt{\frac{C_\calO}{C}} \paren{\cos \phi +  \frac{C' \ell^r \sin \phi}{2 L_z}} \\
     {\WLL} \UDb 2 2 &= \sqrt{\frac{A_\calO B_\calO C_\calO}{ABC}} \paren{\frac{\ell^r_\calO}{\ell^r} + \frac{\ell^r_\calO}{2} \paren{E^2 C' - L_z^2 A'} I_{ABC}} \quad
     {\WLL} \UDb 3 2 = \sqrt{A_\calO B_\calO C_\calO} \frac {C' \ell^r \ell^r_\calO E }{2L_z} I_{ABC} + \frac{\Psi}{Q} {\WLL} \UDb 22 \\ {\WLL} \UDb 2 0 &= \frac{Q E}{2L_z\sqrt{ABC}} \paren{C_\calO I_{ABC} \paren{E^2 C' - L_z^2 A'} + L_z^2 \paren{A' I_{AB} - C' I_{BC}} + \frac{2 \paren{C_\calO - C}}{\ell^r}} - \Psi\\
     {\WLL} \UDb 3 0 &= \frac{Q}{2L_z^2} \paren{C_\calO - C + C' \ell^r (C_\calO E^2 I_{ABC} - L_z^2 I_{BC})} + {\WLL} \UDb 2 0\frac{\Psi}{Q} + \frac{\Psi^2}{2Q}
    \end{split}
\end{equation}

\begin{equation}
\begin{split}
    {\WLX} \UDb 1 1 &= \frac{C' \ell^r - C'_\calO \ell^r_\calO}{2 \sqrt{C C_\calO}} \cos \phi - \sqrt{\frac C {C_\calO}} \paren{\frac{L_z}{C} + \frac{C' \ell^r C'_\calO \ell^r_\calO}{4 C L_z}} \sin \phi \\
    {\WLX} \UDb 2 2 &= \frac{1}{2 \sqrt{A A_\calO B B_\calO C C_\calO}} \paren{\frac{E^2 C' - L_z^2 A'}{\ell^r_\calO} - \frac{E^2 C'_\calO - L_z^2 A'_\calO}{\ell^r} - \frac{I_{ABC}}{2} \paren{E^2 C' - L_z^2 A'} \paren{E^2 C'_\calO - L_z^2 A'_\calO}} \\
    {\WLX} \UDb 2 0 &= \frac{EQ}{4L_z} \frac{C'_\calO \ell^r_\calO}{\sqrt{ABC}} \paren{L_z^2 A' - E^2 C'} I_{ABC} + \frac{E Q}{2 L_z \ell^r} \frac{\paren{C' \ell^r - C'_\calO \ell^r_\calO}}{\sqrt{ABC}} \\
    {\WLX} \UDb 3 0 &= \frac{Q}{2 L_z^2} \paren{C' \ell^r - C'_\calO \ell^r_\calO} - I_{ABC} C'\ell^r C'_\calO \ell^r_\calO \frac{E^2 Q}{4 L_z^2} + \frac \Psi Q {\WLX} \UDb 2 0
    \end{split}
\end{equation}    

Note that these relations take the simplest form in the intermediate SNT, as defined in Section \ref{sec:parallel}, in which we simply have $\Psi=0$.

\bibliography{main}

\end{document}